\documentclass[twocolumn]{aastex63}

\usepackage{xcolor}

\usepackage{xspace}
\usepackage{amsmath}

\mathchardef\mhyphen="2D

\begin{document}
%% Front Matter
\title{A Criterion for the Onset of Chaos in Compact, Eccentric Multiplanet Systems}

\author[0000-0002-9908-8705]{Daniel Tamayo}\thanks{NHFP Sagan Fellow: dtamayo@astro.princeton.edu}
\affiliation{Department of Astrophysical Sciences, Princeton University, 4 Ivy Lane, Princeton, NJ 08540, USA}
\author{Norman Murray}
\affiliation{Canadian Institute for Theoretical Astrophysics, University of Toronto, 60 St. George Street, Toronto, ON M5S 3H8, Canada}
\author{Scott Tremaine}
\affiliation{Canadian Institute for Theoretical Astrophysics, University of Toronto, 60 St. George Street, Toronto, ON M5S 3H8, Canada}
\affiliation{Institute for Advanced Study, Einstein Drive, Princeton, NJ 08540, USA}
\author{Joshua Winn}
\affiliation{Department of Astrophysical Sciences, Princeton University, 4 Ivy Lane, Princeton, NJ 08540, USA}

\begin{abstract}
We derive a semi-analytic criterion for the presence of chaos in compact, eccentric multiplanet systems. Beyond a minimum semimajor-axis separation, below which the dynamics are chaotic at all eccentricities, we show that (i) the onset of chaos is determined by the overlap of two-body mean motion resonances (MMRs), like it is in two-planet systems; (ii) secular evolution causes the MMR widths to expand and contract adiabatically, so that the chaotic boundary is established where MMRs overlap at their greatest width. 
For closely spaced two-planet systems, a near-symmetry strongly suppresses this secular modulation, explaining why the chaotic boundaries for two-planet systems are qualitatively different from cases with more than two planets.
We use these results to derive an improved angular-momentum-deficit (AMD) stability criterion, i.e., the critical system AMD below which stability should be guaranteed.
This introduces an additional factor to the expression from \cite{Laskar17} that is exponential in the interplanetary separations, which corrects the AMD threshold toward lower eccentricities by a factor of several for tightly packed configurations.
We make routines for evaluating the chaotic boundary available to the community through the open-source SPOCK package.
~\\ % this is here to put some space between the abstract and Section 1
\end{abstract}

\section{Introduction}

Radial-velocity surveys \citep[e.g.,][]{Udry19} and transit surveys \citep{Lissauer11} have both revealed a substantial population of compact multiplanet systems.
Various authors have defined the term ``compact'' in different ways \citep[e.g.,][]{Lissauer11, Volk15}.
We will use a dynamically motivated definition: a system (or subsystem) of planets is compact if the period ratio of any adjacent pair of planets is between 1 and $2$.
Within this period-ratio range, first-order mean-motion resonances (hereafter MMRs), with period ratios of $j:j-1$, strongly modify the dynamical structure of phase space; at wider separations (and low orbital eccentricities), the strongest resonances are typically secular (between the orbits' apsidal and nodal precession frequencies). The restriction in the range of period ratios also allows us to simplify many algebraic expressions using the approximation $|\Delta a|/a \ll 1$,
where $a$ refers to the semimajor axis.
By this definition, Venus, Earth and Mars are a compact trio, and Uranus and Neptune are a compact pair.
In the exoplanet context, $\approx\,40\%$ of 3+ planet systems have at least one pair of adjacent planets that is compact.

Exoplanets typically complete $10^{10}\mhyphen 10^{12}$ orbits between the time they form and the time we observe them.
Given these vast spans of time, a central question for connecting observations to planet-formation models is whether planetary populations look the same after 10\,Gyr as they do following disk dispersal at ages of a few Myr. 
Several studies have used N-body experiments to argue that multiplanet systems evolve significantly over their lifetimes, starting with larger numbers of planets and subsequently destabilizing and colliding to form progressively longer and longer lived configurations with fewer and fewer planets \citep[e.g.,][]{Pu15, Volk15}.
However, our theoretical understanding of the dynamics driving these instabilities remains incomplete.

The ability to predict the stability of arbitrary orbital configurations would also help to constrain uncertain planetary masses and orbital parameters.
This argument is based on the ``temporal Copernican principle'' --- a corollary of which is that an observer examining a planetary system at an arbitrary moment in the system's lifetime should not expect to detect it just prior to a major cataclysm \citep[e.g.,][]{Gott93}. In other words, one would expect a randomly discovered system to have comparable past and future lifetimes.
One can therefore reject values of planet masses and orbital parameters for which the system's estimated future lifetime would be much shorter than its current age, making it possible to constrain planetary masses and orbital parameters that are challenging to measure directly \citep{Tamayo21}. 
This approach has occasionally been implemented by directly checking for stability over timescales of $\sim10^9-10^{10}$ orbits through brute-force numerical integrations for some of the most important exoplanet discoveries \citep[e.g.,][]{Steffen13, Tamayo15, Tamayo17, Quarles17, Wang18, Rosenthal19}. 
However, the Gyr ages of typical systems mean that $N$-body integrations are orders of magnitude too computationally expensive for a full exploration of parameter space, even for a single system.

Many authors, starting with \cite{Chambers96}, have used N-body integrations to explore the dependence of instability times on the initial conditions and parameters of multiplanet systems \citep{Yoshinaga99, Marzari02, Faber07, Zhou07, Smith09, Funk10, Pu15, Obertas17, Gratia19, Lissauer21}.
While these numerical experiments have provided insight into the dynamical mechanisms that drive these instabilities, the practical restrictions that were variously imposed on the allowed parameter space (e.g., equal spacing, equal orbital eccentricities, equal masses, etc.) lead to inaccurate instability time predictions for real systems, often by orders of magnitude \citep{Cranmer21}.

\cite{Tamayo16} and \cite{Tamayo20} tackled this high-dimensional prediction problem for compact multiplanet systems using machine learning techniques.
They demonstrated that their Stability of Planetary Orbital Configurations Klassifier (SPOCK) can accurately assess the stability of compact systems over $10^9$ orbits based on an analysis of a much shorter integration.
This results in computation times up to $10^5$ times faster than direct N-body integrations.
Recently, \cite{Cranmer21} generalized this result with a deep learning model that can not only classify a system's stability, but also predict the time to instability out to $10^9$ orbits. 

In this paper we pursue the problem of predicting the long-term stability of compact planetary systems analytically. 
Such efforts also have a long history, and it has become clear that distinct dynamical mechanisms limit the stability of planetary systems in different regions of parameter space.
In order to put our work into context and to introduce several of the concepts and results we will use below, we begin by summarizing some of the previous key results.

\section{Background}

A central challenge in understanding the stability of compact $N$-planet system is that there are $7N$ parameters to consider (6 orbital elements and 1 mass per planet).
Analytical work has therefore progressed by starting from simple cases and adding parameters sequentially.
One powerful simplification we will use throughout this paper is to consider only the planar case.
As we motivate numerically below, this is an excellent approximation for mutual inclinations less than a few degrees, typical of transiting multiplanet systems \citep[e.g.,][]{Fabrycky14}.

\subsection{Two-Planet Systems}

The first analytic investigations considered the case of two bodies on initially circular orbits.
\cite{Wisdom80} showed that, at the smallest separations, chaos is driven by the overlap of first-order ($j:j-1$) MMRs\footnote{Throughout this paper, we will use the term `order' to describe the power of the eccentricity $e$ carried by a particular term. The strength of a $j:j-k$ two-body MMR is proportional to $e^k$ \citep[e.g.,][]{Murray99}. First-order ($k=1$) MMRs are thus the strongest two-body MMRs for small eccentricities.}.
The intervals between the period ratios of adjacent first-order MMRs become smaller and smaller as $j$ increases (the 2:1 is further in period ratio from the 3:2 than the 3:2 is from the 4:3, etc.).
At the same time, the widths of these ever more tightly spaced resonances increase due to the decreasing interplanetary separation (even at zero eccentricity). \cite{Wisdom80} showed that this implies a threshold interplanetary separation below which the MMR widths overlap, leading to chaotic motion,
\begin{equation}
\Bigg(\frac{a_2 - a_1}{a_1}\Bigg)_{\rm{\!\!crit}} \approx 1.46\,\mu^{2/7}, \label{wisdom}
\end{equation}
where the $a_i$ are the semimajor axes, and $\mu$ is the planet-star mass ratio.
This critical spacing, originally derived in the restricted case of a planet, star, and test particle, can be generalized to the case of two planets with non-zero masses by replacing the mass ratio $\mu$ by the sum of the planetary mass ratios \citep{Deck13},
\begin{equation}
\mu = \frac{m_1 + m_2}{M_\star}, \label{mu}
\end{equation}
where the $m_i$ represent the planetary masses, and $M_\star$ the stellar mass.

\cite{Wisdom80} also showed that MMR widths grow with increasing orbital eccentricities.
Thus, for given masses and separations beyond the critical separation for cirular orbits (Eq.\:\ref{wisdom}), there exists a critical eccentricity beyond which first-order MMRs will overlap and drive chaos. 
However, while higher order MMRs have negligible widths in the initially circular case \citep{Ferraz07} and thus can be ignored, \cite{Hadden18} show that they must be considered at higher eccentricities where their widths can be substantial.

The results of \cite{Hadden18} are based on two key developments that facilitate a simple analytic criterion for resonance overlap.

\paragraph{The relative eccentricity} It has long been known that  the dynamical evolution near a first-order MMR does not depend on both planets' eccentricity vectors independently, but only on a particular linear combination of the two \citep{Sessin84, Wisdom86}. \cite{Hadden19} showed that this result also holds to excellent approximation for higher order MMRs in the compact limit $|\Delta a/a|\ll1$.

In particular, at the closest separations---where the equations of motion reduce to Hill's equations \citep{Henon86}---there are two additional conserved  quantities \citep[e.g.,][]{Goldreich81}, which at low eccentricities can be thought of as a ``center-of-mass" eccentricity vector. 
Defining eccentricity vectors in the complex plane $\boldsymbol{e}_i = e_i \exp{(i\varpi_i})$ with magnitude given by the eccentricity $e_i$ and direction given by the longitude of pericenter $\varpi_i$ (henceforth vectors will be denoted in boldface), the additional conserved quantity in the Hill problem is, to leading order in the eccentricities, $\boldsymbol{e_{com}} = (m_1 \boldsymbol{e_1} + m_2 \boldsymbol{e_2})/(m_1+m_2)$.
Therefore, in the Hill problem the dynamics do not depend on the two eccentricity vectors independently but rather only on their vector difference $\boldsymbol{e_{12}} \equiv \boldsymbol{e_2}-\boldsymbol{e_1} \equiv e_{12}\exp(i\varpi_{12}).$ \citep{Henon86}.

The expansion of the disturbing function of the planet interaction potential includes several ``sub-resonances'' corresponding to a $j:j-k$ MMR, which are proportional to $e_1^{l}e_2^{k-l} \cos\left[j\lambda_2 - (j-k)\lambda_1-(k-l)\varpi_2 - l\varpi_1\right]$ for $l=0,\dots,k$.
As argued by \cite{Hadden19}, the fact that in the Hill limit the disturbing function can only depend on the eccentricities through the combinations $e_{12}$ and $\varpi_{12}$ implies that the sum over these sub-resonances must be expressible as a single resonant term proportional to $e_{12}^k \cos\left[j\lambda_2 - (j-k)\lambda_1-k\varpi_{12}\right]$.

While these results strictly apply only in the Hill limit, \cite{Hadden19} showed that they persist to a good approximation\footnote{\cite{Hadden19} derive slightly modified definitions of the center-of-mass eccentricity and eccentricity difference that depend on the interplanetary separation and provide an even better approximation. However, in order to combine the widths of different MMRs, which lie at different interplanetary separations, we follow \cite{Hadden18} in ignoring these corrections.} far beyond the regime where Hill's equations apply, and in fact hold reasonably well for all period ratios $\lesssim 2$.

\paragraph{The orbit-crossing eccentricity} In general, MMR strengths depend on both the interplanetary separation and the orbital eccentricities. In a second important development, \cite{Hadden18} showed that in the compact limit ($|\Delta a/a| \ll 1$), the dependence on the semimajor axes can be absorbed by expressing eccentricities as a fraction of their orbit-crossing value $e_{12}^{\rm{cross}}$ 
\begin{equation}
e_{12}^{\rm{cross}} = \frac{a_2 - a_1}{a_2} = 1 - \Bigg(\frac{j-k}{j}\Bigg)^{\!\!2/3} \approx \frac{2k}{3j}, \label{ecross}
\end{equation}
We therefore define the normalized, relative eccentricity vector\footnote{This is equivalent to $\mathcal{Z}_{12}/\mathcal{Z}_{12}^{\rm{cross}}$ in the notation of \cite{Hadden18}, though we choose $e_{12} = \sqrt{2}\mathcal{Z}_{12}$ and $e_{12}^{\rm{cross}} = \sqrt{2} \mathcal{Z}_{12}^{\rm{cross}}$ for simplicity of notation and geometric interpretation.}
\begin{equation}
\boldsymbol{\tilde{e}_{12}} \equiv \frac{\boldsymbol{e_2} - \boldsymbol{e_1}}{e_{12}^{\rm{cross}}}. \label{etilde}
\end{equation}
With this definition, $\boldsymbol{\tilde{e}_{12}} = 1$ is the collision criterion for two arbitrarily oriented, coplanar orbits in the compact limit \citep{Kholshevnikov99, Laskar17}.

In the pendulum approximation, which is valid if the eccentricity is not too close to zero \citep[e.g.,][]{Murray99},  the half-width $\delta a$ of a $k^\mathrm{th}$ order MMR is \citep{Hadden18} 
\begin{equation}
\frac{\delta a}{a} \approx 4 \sqrt{\frac{h_k}{3}} \mu^{\frac{1}{2}} \tilde{e}^{\frac{k}{2}}, \label{MMRwidth}
\end{equation}
plus terms of higher order in the eccentricity. 
Here $h_k$ is a constant specific to each set of $k$th order MMRs  (and is of order unity for modest values of $k$), and $\mu$ and $\tilde{e}$ are defined in Eqs.\:\ref{mu} and \ref{etilde}, respectively.

\begin{figure*}
    \resizebox{0.99\textwidth}{!}{\includegraphics{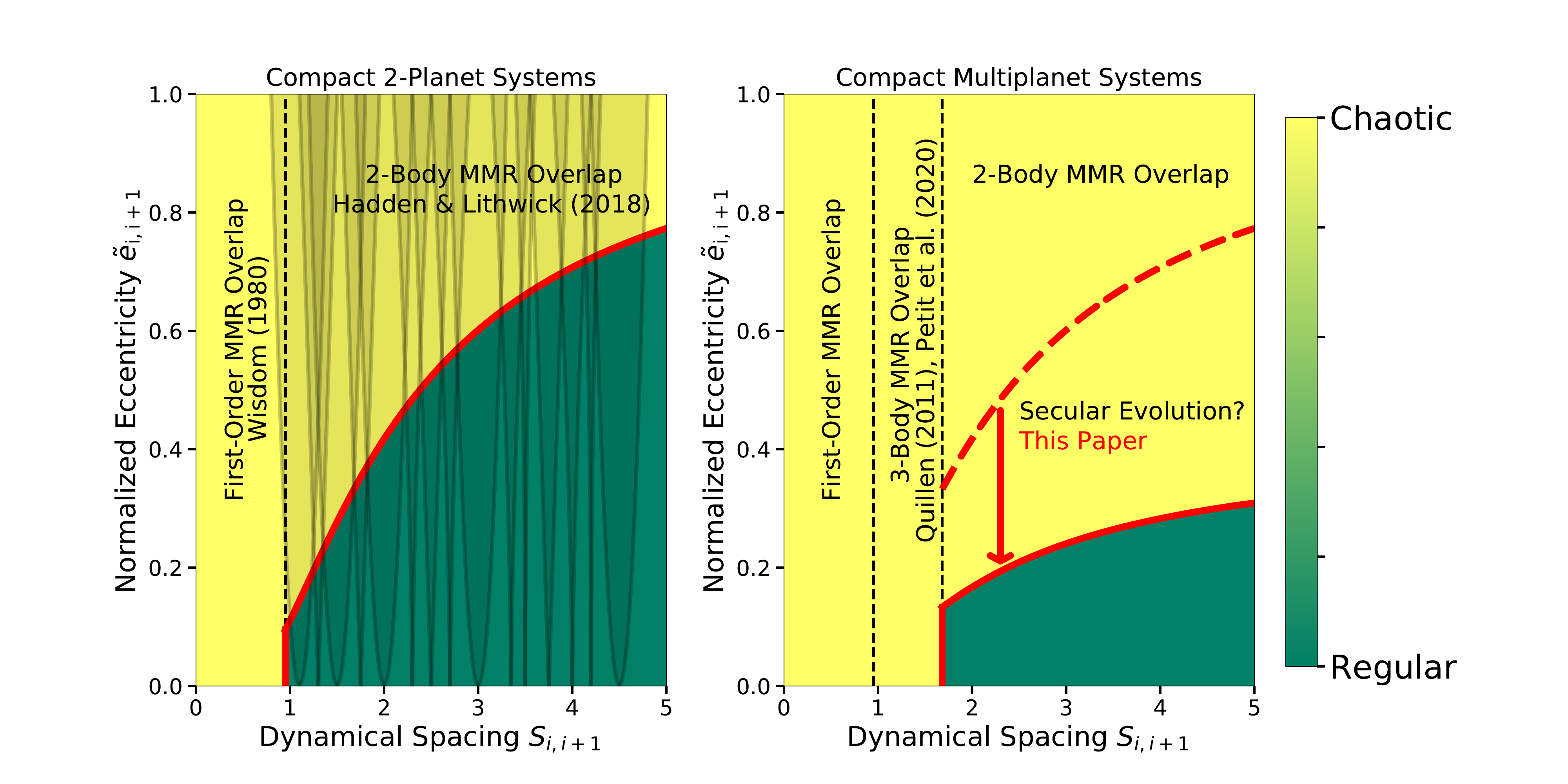}}
    \caption{Different chaotic boundaries imposed by various dynamical mechanisms, and studies deriving those criteria. Yellow regions are chaotic, while green regions are not The $x$-axis is the dynamical spacing between adjacent planets (Eq.\:\ref{S}), while the $y$-axis is the linear combination of eccentricity vectors that dominantly sets the resonant dynamics (Eq.\:\ref{etilde}). The left panel shows the two-planet case, with schematic MMR widths that start overlapping at the red boundary (Eq.\:\ref{H18crit}). The right panel shows an equal-mass, equal-spacing, equal-eccentricity multiplanet case, where 3-body MMRs and secular evolution additionally can play a major role.
    \label{fig:schematic}}
\end{figure*}

\cite{Hadden18} consider the aggregate sum of all MMR widths, and calculate a covering fraction or optical depth $\tau$ of MMRs within a fixed range of interplanetary separations.
A key parameter, which we will call the ``dynamical separation" $S_{12}$ between a pair of planets, is given by
\begin{equation}
S_{12} = \frac{a_2-a_1}{R_{1/4}},\: {\rm where}\: R_{1/4} \equiv a_2\,\mu^{1/4} \label{S}
\end{equation}
and $\mu$ is defined by Equation (\ref{mu})\footnote{Another way to portray $S$ is as a normalized crossing eccentricity, $e^{\rm{cross}}/\mu^{1/4}$. We note that $S$ does not increase without bound for wide interplanetary separations, but instead approaches a constant value of $\mu^{-1/4}$.}. They derive a covering fraction of two-body MMRs between the pair of planets,
\begin{equation}
\tau_{12} \approx \Bigg(\frac{1.8}{S_{12}}\Bigg)^{\!2} |\ln (\tilde{e}_{12})|^{-3/2}. \label{tauH18}
\end{equation}
Solving for the critical eccentricity at which $\tau$ equals unity and MMRs are overlapped, \cite{Hadden18} obtain a chaotic boundary for two planets
\begin{equation}
\tilde{e}_{12}^{\rm{crit}} \approx \exp{\Bigg[-\Bigg(\frac{1.8}{S_{12}}\Bigg)^{\!\!4/3\,}\Bigg]}\;\;(2\:\rm{planets}). \label{H18crit}
\end{equation}
In the left panel of Fig.\:\ref{fig:schematic}, we schematically plot the widths of various MMRs in gray, which grow with increasing eccentricity.
The boundary at which the covering fraction of MMRs reaches unity is plotted in red \citep{Hadden18}.
In summary, the chaos in two-planet systems is always driven by the overlap of two-body MMRs.
At the closest separations, first-order MMRs overlap even at zero eccentricity \citep{Wisdom80}.
At wider separations, MMRs only overlap at non-zero eccentricity, and higher order MMRs need to be considered \citep{Hadden18}.

We can also gain some intuition from the following approximations (discussed and compared to the above expressions in Appendix \ref{sec:critcomp}), which are valid at tight separations where $S \lesssim 1.8$,
\begin{eqnarray}
\tau_{12} &=& \Bigg(\frac{2.9}{S_{12}}\Bigg)^{\!\!2}\tilde{e}_{12} \:\:\:\:\:\:\:\:\:\:\:\:\:\: (\tilde{e} \lesssim 0.4) \nonumber \\
\tilde{e}_{12}^{\rm{crit}} &=& \Bigg(\frac{S_{12}}{2.9}\Bigg)^{\!\!2} \:\:\:\:\:\:\:\:\:\:\:\:\:\:\:\:\:\:\:\: (\tilde{e} \lesssim 0.4), \label{lowe}
\end{eqnarray}
From these equations and Eq.\:\ref{S}, we see that $\tau$ scales approximately quadratically with the fractional interplanetary separation, linearly with the eccentricity, and as the square root of the mass ratio (and has a steeper scaling with eccentricity for $\tilde{e} > 0.4$).
Because the stability criterion depends more strongly on eccentricity than mass ratio, when the requirement of long-term stability is applied to exoplanet systems, the resulting eccentricity constraints are typically stronger than the constraints on masses \citep{Tamayo21}.

In addition, the three-body problem allows for a rigorous criterion whereby close encounters are forbidden beyond a critical value of a dimensionless parameter combining the total angular momentum and energy \citep{Marchal82}.
This general result was applied to planetary systems by \cite{Gladman93}, where it is commonly referred to as the Hill criterion.
At low eccentricities (see \citealt{Petit18} for the eccentric two-planet case), close encounters are forbidden for interplanetary separations $|\Delta a| > 2\sqrt{3}\,R_H$, where $R_H$ denotes the mutual Hill radius $R_H = a_1 (\mu/3)^{1/3}$.
Partly because of the Hill criterion,  dynamical separations in compact  systems are traditionally expressed as a multiple of the mutual Hill radius.
However, the Hill criterion is not valid for systems with more than two planets, while resonance overlap criteria can be generalized to higher multiplicities.

We may therefore expect $S$ (Eq.\:\ref{S}) to be a more useful measure of dynamical separation than the Hill separation in multiplanet systems.
In practice, the differences between the scaling of the Hill criterion, $|\Delta a|\sim \mu^{1/3}$, the resonance-overlap criterion for circular orbits, $|\Delta a|\sim \mu^{2/7}$, and the dynamical separation, $|\Delta a|\sim \mu^{1/4}$, are small enough to be challenging to distinguish in $N$-body experiments.
However, comparing the ends of the planet-mass distribution, the differences become significant.
A value of $S=2.5$ that limits $\tilde{e}$ to be below $\approx 0.5$ (Eq.\:\ref{H18crit}) corresponds to a Hill separation of $\approx 10 R_H$ for a pair of Earth-mass planets around the Sun, while it translates to $\approx 6 R_H$ for two Jupiter-mass planets.

\subsection{Initially Circular Multiplanet Systems}

The instability boundary in two-planet systems is sharp, with configurations largely divisible into those that go unstable on timescales comparable to the synodic period between planet conjunctions, and those that remain stable over typical planetary system lifetimes \citep{Gladman93}.
By contrast, a large number of numerical results starting with those of \cite{Chambers96} have shown that initially circular systems of three or more planets exhibit a much broader dynamic range of instability timescales.

\cite{Quillen11}, and more recently \cite{Petit20}, explain this behavior as a consequence of the overlap of three-body resonances.
In the range of period ratios between first and second order MMRs (these MMRs have finite widths at zero eccentricity and such configurations often exhibit much shorter instability times in N-body integrations, e.g., \citealt{Obertas17}), this is a compelling explanation. 
Both secular resonances and higher order MMRs have vanishing widths at zero eccentricity \citep{Ferraz07} --- and therefore the overlap of those resonances cannot be driving the instabilities observed in numerical integrations.

So far we have been considering a perturbative expansion in powers of the (typically small) eccentricities, but at leading (first) order in the mass ratio $\mu$.
In this limit, the strongest resonances are first-order two-body MMRs, with strengths $\propto \mu \tilde{e}$ \citep[e.g.,][]{Murray99} and widths $\propto \mu^{1/2} \tilde{e}^{1/2}$ (Eq.\:\ref{MMRwidth}).
At the next power in the masses $\propto \mu^2$, however, there appear three-body MMRs involving the orbital parameters of trios of planets.
These can again be separated into three-body MMRs of different orders in the eccentricity.
Unlike the case of two-body MMRs (excluding co-orbital resonances), there exist three-body MMRs that are zeroth-order in (i.e., independent of) the eccentricity \citep[e.g.,][]{Quillen11}.
While these three-body MMRs are therefore weaker by a factor of $\mu$ than two-body MMRs, their independence of the eccentricity renders them important for circular orbits.
Indeed, \cite{Quillen11} showed that the finite widths of such three-body MMRs at zero eccentricity allow them to overlap and drive chaos at interplanetary separations that are significantly wider than the two-planet criterion (Eq.\:\ref{wisdom}) of \cite{Wisdom80}.
Building on that work by accounting for variable chaotic diffusion rates in different regions of the three-body resonance network, \cite{Petit20} were able to accurately predict the instability times measured in numerical integrations of such systems.

While the instability times in regions where three-body resonances overlap span a broad range of timescales, they are almost always smaller than the typical main-sequence lifetimes ($\sim\,10^{11}$ orbits) of the known stars hosting compact multiplanet systems. 
The critical separation inside which three-body resonances overlap therefore provides a sufficient criterion for instability in typical planetary systems.
\cite{Petit20} derive this critical separation for the general case in which the masses and interplanetary spacings between a trio of planets are unequal.
For the simple case of equal masses and separations (EMS), the results of \cite{Quillen11} and \cite{Petit20} can be written as\footnote{There is a typographical error in Eq.\ 61 of \citealt{Petit20} (A.\ Petit, personal communication).}
\begin{equation}
S_{12\rm{crit}} \approx 1.68\:\:(\rm{3\mhyphen body\:MMR\:overlap,\:EMS}). \label{3bMMRcrit}
\end{equation}
In the case of EMS systems, one can write the 3-body MMR overlap criterion in terms of the dynamical separation $S$ between adjacent planets (with $\mu$ given by twice the planet-star mass ratio through Eq.\:\ref{mu}), but in the general case the criterion depends on both of the separations between adjacent planets, and $\mu$ is replaced by a generalized mass parameter which is not simply the sum of the mass ratios \citep[see Eqs.\ 45 and 53 of][]{Petit20}.

\begin{figure}
    \centering \resizebox{\columnwidth}{!}{\includegraphics{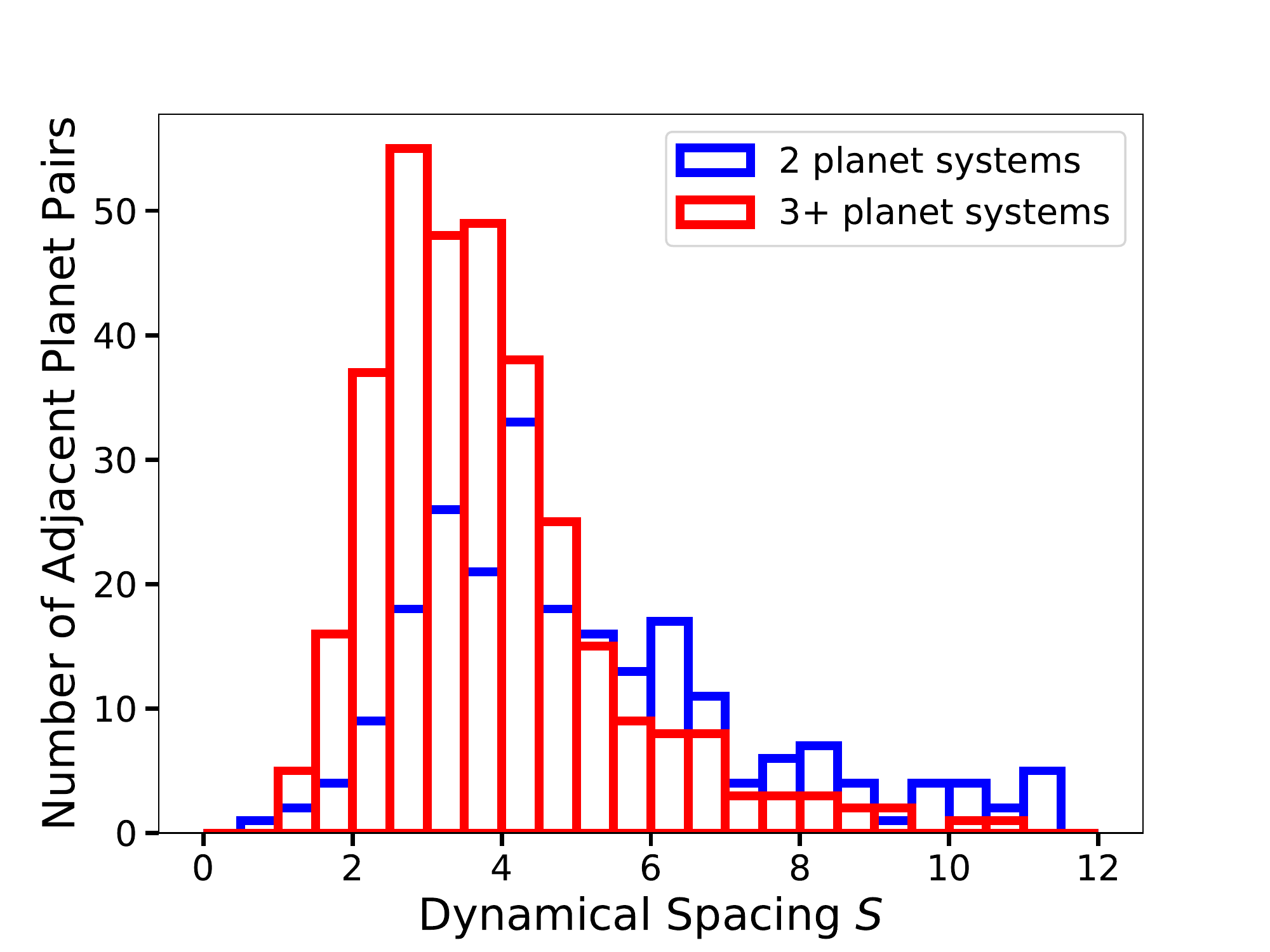}}
    \caption{Distribution of dynamical spacings (Eq.\:\ref{S}) from the transiting multiplanet sample of \cite{Weiss18}, with masses estimated using a mass-radius relationship. Histograms are separated into systems with two known planets (blue) and those with three or more known planets (red).
    \label{fig:spacing}}
\end{figure}

Figure \ref{fig:spacing} shows the distribution of dynamical spacings between adjacent planets for the 909 planets in 355 multi-planet systems analyzed by \cite{Weiss18} as part of the California Kepler Survey \citep{Petigura17}.
Calculating these spacings requires adopting planet masses, which \cite{Weiss18} estimate using the mass-radius relations from \cite{Weiss14}. 
We plot separate histograms for systems with two known planets and ones with higher planet multiplicities.
Fewer than 2\% of adjacent planet pairs in 3+ planet systems have $S < 1.68$.
More careful analysis of these most closely packed systems using the general expressions of \cite{Petit20} for unequal spacings and masses reveal that three-body resonances do not overlap in these systems.

This is unsurprising given that such configurations would have destabilized and rearranged long ago.
However, the observed distribution of dynamical spacings seems suppressed even beyond $S=1.68$, perhaps hinting that many more widely spaced configurations can nevertheless be unstable.
In particular, while initially circular orbits may be stable beyond $S=1.68$, eccentric orbits need not be.
This expectation is consistent with numerical experiments showing that stability is a strong function of eccentricities \citep[e.g.,][]{Zhou07, Gratia19}.

This motivates trying to understand the general case, particularly given that we expect planets to have eccentric orbits.
Terrestrial planet formation models predict eccentric orbits \citep[e.g.,][]{Hansen12, Dawson16, Tremaine15}, and transit timing variation \citep{Lithwick12, Jontof14, Hadden16, Hadden17} and transit duration measurements \citep{vanEylen15, Xie16, vanEylen19} confirm this expectation.

\subsection{Eccentric Multiplanet Systems (this paper)}

At the outset of this work, it was not
clear to us whether the stability of most multiplanet configurations at wider separations remains limited by the overlap of two-body MMRs, as in the two-planet case, or by eccentricity-dependent, higher order 3-body MMRs.
A clue to the underlying dynamics is that the machine learning model \texttt{SPOCK} \citep{Tamayo20} was only trained on three-planet systems in which one pair was in (or near) a strong 2-body MMR, and yet the trained model made accurate predictions even for randomly sampled compact systems.
As argued by \cite{Tamayo20}, this suggests that the chaotic dynamics in eccentric compact systems is dominated by two-body MMRs and not 3-body MMRs.

We took an approach that is qualitatively similar to that of \cite{Petit17}, who exploited the wide separation of timescales between the fast resonant dynamics and the much slower secular modulation of the eccentricities $\tilde{e}$.
In particular, this slow modulation causes the eccentricity-dependent MMR widths (Eq.\:\ref{MMRwidth}) to adiabatically grow and shrink over a secular cycle \citep[e.g.,][]{Wisdom85}, which in turn affects whether or not MMRs overlap.
However, our work differs in three main ways.

The first and simplest difference is that we consider the secular modulation of a different MMR boundary.
\cite{Petit17} consider whether the secular evolution can raise the eccentricities to the point at which first-order MMRs overlap (or to $\approx\,e_{\rm{cross}}$ in cases where the latter is smaller; \citealt{Laskar17}).
However, \cite{Hadden18} showed that the overlap of MMRs of all orders provides a more accurate match to the chaotic boundary.
We therefore instead consider the modulation of the two-planet boundary of \cite{Hadden18}, which, as we will see below, often yields chaotic boundaries that are a factor of two or more lower in the eccentricities than the estimates of \cite{Petit17} for compact systems.

Second, \cite{Petit17} and \cite{Laskar17} consider the long-term redistribution of the system's angular momentum deficit \citep[AMD;][]{Laskar00}
\begin{equation}
{\rm AMD} = \sum_{j=1}^N m_j \sqrt{GM_\star a_j}\Bigg(1-\sqrt{1-e_j^2}\cos{i_j}\Bigg), \label{AMD}
\end{equation}
where $G$ is the gravitational constant, the $m_j$ are the planetary masses, the $e_j$ the orbital eccentricities, the $i_j$ the orbital inclinations, and the sum runs over all $N$ planets.
In particular, they consider the worst-case scenario, in which the system exhibits secular chaos which transfers all of the system's AMD to the $\tilde{e}$ of a single pair of planets.
The advantage of this estimate based on the worst-case scenario is simplicity: for given planetary masses and semimajor axes, it generates a single system-wide critical AMD value, below which the system should always be stable. 
However, it is not clear how to interpret the criterion in the limit in which one body is massless (in which case its eccentricity would always be able to reach unity) or in the limit of a distant planetary or stellar companion (which can raise the AMD to arbitrarily large values and would allow the eccentricities of all the inner bodies to reach unity).

Additionally, if most compact systems were {\it not} significantly affected by secular chaos, then the threshold eccentricities that guarantee AMD stability, while technically correct, would be too low, hedging against a possibility that seldom occurs in practice.
While secular chaos in our more widely separated solar system drives major changes in Mercury's orbit on timescales of $10^{10}$--$10^{11}$ orbits \citep{Laskar09}, and secular chaos has been demonstrated for some  widely separated giant planets \citep{Wu11} as well as some compact multiplanet configurations \citep{Volk20}, it is also plausible that for most low-mass compact multiplanet systems, the requirement to avoid 2-body MMR overlap constrains the eccentricities to low enough values that secular resonances do not overlap and secular chaos is not important.

Rather than trying to address this difficult question, we restrict ourselves in this paper to deriving a chaos criterion for two-body MMR overlap in compact multiplanet systems in the presence of eccentricity modulation by the shorter-term {\it regular}, i.e., non-chaotic, secular dynamics.
This will allow us to check our derived boundaries against the results from computationally tractable $N$-body integrations and to establish whether the shortest chaotic timescales are set by two-body MMR overlap, or whether other resonances, e.g., higher order 3-body MMRs, need to be considered.

The third difference between the present work and \cite{Petit17} and \cite{Laskar17} is that we consider not only the most tightly packed planet pair but all of the planet pairs. We show that in compact systems it is necessary to consider two-body MMRs between different pairs of planets in the system simultaneously.

We begin in Sec.\:\ref{sec:2p} by exploring the secular modulation of $\tilde{e}$ in the two-planet case, and showing why this effect is negligible in the case studied by \cite{Hadden18}. 
We consider multiplanet systems in Sec.\:\ref{sec:multi}, first analyzing the effect of two-body MMRs with multiple bodies in the system (Sec.\:\ref{sec:MMRs}), and then isolating the effects of the secular dynamics (Sec.\:\ref{sec:secular}). 
We put everything together in Sec.\:\ref{sec:disc}, and conclude in Sec.\:\ref{sec:conc}.

\section{Two Planets} \label{sec:2p}
To approach the problem analytically, we exploit the hierarchy of timescales in the problem.
Because the secular timescales are significantly longer than both the orbital periods and the MMR libration timescales, we average over the latter two fast timescales to isolate the secular evolution.
To lowest order in the eccentricities, this is the classical Laplace-Lagrange problem, in which the semimajor axes remain constant.
While we ignore MMR contributions to the precession rates, which can be significant near and inside MMRs \citep{Wisdom85, Sansottera19}, we will find that this simplest approximation is sufficient in most cases for determining the chaotic boundary.

The planar Laplace-Lagrange problem with two planets has four dynamical variables corresponding to each planet's orbital eccentricity and pericenter orientation, with the total energy and angular momentum as conserved quantities. Thus the two-planet Laplace-Lagrange system has two phase-space dimensions or one degree of freedom. In the compact limit, there is an additional conserved quantity $\boldsymbol{e_{com}}$, which removes the remaining freedom for the orbits to exchange angular momentum, so the eccentricities remain fixed (i.e., one of the Laplace-Lagrange eigenfrequencies goes to zero).
While this conserved quantity only exists in the limit of the Hill problem, i.e., at the closest separations, the vestiges of this symmetry persist across the period ratio range where strong MMRs exist ($P_2/P_1 \lesssim 2$), strongly suppressing secular eccentricity oscillations --- as we show below.  We argue that this is one of the fundamental reasons why the chaotic boundaries for two-planet and three-planet systems are qualitatively different from one another.

\subsection{Coordinates and Notation}

We develop our Hamiltonian treatment in canonical Poincaré variables, using democratic heliocentric coordinates \citep[see][]{Duncan98, Hernandez17}. 
In that case the actions and their conjugate angles are 
\begin{eqnarray}
\Lambda_{i} = m_i \sqrt{GM_\star a_{i}} &\leftrightarrow& \lambda_i \\
\Gamma_i = \Lambda_{i} \Bigg(1 - \sqrt{1-e_i^2}\Bigg) &\leftrightarrow& \gamma_i = -\varpi_i \label{canpoincare}
\end{eqnarray}
where the $\lambda_i$ are the mean longitudes, and the $\varpi_i$ are the longitudes of pericenter.
Because we average over the fast MMR oscillations the problem is purely secular, we may assume that the semimajor axes and their associated actions $\Lambda_i$ remain fixed, while the $\Gamma_i$ evolve with time.
We also define complex variables 
\begin{equation}
\boldsymbol{G_i} \equiv \sqrt{2\Gamma_i} e^{i\gamma_i}. \label{Gi}
\end{equation}

\subsection{Laplace-Lagrange Hamiltonian} \label{sec:laplag}

The secular Hamiltonian for two planets is 
\begin{equation}
\mathcal{H} = -\frac{Gm_1m_2}{a_{20}}\mathcal{R}_{sec},
\end{equation}
where, to leading order in the eccentricities \citep[e.g.,][]{Murray99},
\begin{equation}
\mathcal{R}_{sec} = \frac{\alpha}{8}\Bigg[b_{3/2}^1(\alpha)(e_1^2 + e_2^2) -2b_{3/2}^2(\alpha)e_1e_2\cos(\varpi_2 - \varpi_1)\Bigg],
\end{equation}
where $\alpha = a_{1}/a_{2}$ is the semimajor axis ratio, and the $b_s^j$ are Laplace coefficients evaluated at $\alpha$.

The secular disturbing function can be written in matrix notation in terms of the complex eccentricity vectors $\boldsymbol{e_i}$ as \citep[e.g.,][]{Hadden19},
\begin{equation}
\mathcal{R}_{sec} = \frac{\alpha}{8} \begin{pmatrix} \boldsymbol{e_1}^* & \boldsymbol{e_2}^* \end{pmatrix}
\begin{pmatrix}
b_{3/2}^{1}(\alpha) & -b_{3/2}^{2}(\alpha) \\
-b_{3/2}^{2}(\alpha) & b_{3/2}^{1}(\alpha)
\end{pmatrix}
\begin{pmatrix}\boldsymbol{e_1} \\ \boldsymbol{e_2}\end{pmatrix}. \label{Rsec}
\end{equation}
For $e_1,e_2\ll1$ the eccentricity vectors are connected to the canonical variables through
\begin{equation}
\begin{pmatrix}\boldsymbol{e_1}^* \\ \boldsymbol{e_2}^*\end{pmatrix} \simeq \begin{pmatrix}
\sqrt{\frac{1}{\Lambda_{1}}} & 0 \\
0 & \sqrt{\frac{1}{\Lambda_{2}}}
\end{pmatrix}
\begin{pmatrix}\boldsymbol{G_1} \\ \boldsymbol{G_2}\end{pmatrix}. \label{scaling}
\end{equation}

Plugging Eq.\:\ref{scaling} into Eq.\:\ref{Rsec} yields the secular Hamiltonian
\begin{equation}
\mathcal{H} = \frac{1}{2}\begin{pmatrix}\boldsymbol{G_1} & \boldsymbol{G_2}\end{pmatrix}
 \cdot{\mathcal{M}}\cdot
\begin{pmatrix}\boldsymbol{G_1^*} \\ \boldsymbol{G_2^*}\end{pmatrix},
\end{equation}
where, after some algebra,
\begin{equation}
\mathcal{M} = -\frac{n_{2}}{4M_\star} \begin{pmatrix}
\alpha^{1/2}\,b_{3/2}^{1}(\alpha)\,m_2 & -\alpha^{3/4}\, b_{3/2}^{2}(\alpha)\,\sqrt{m_1 m_2} \\
-\alpha^{3/4}\,b_{3/2}^{2}(\alpha)\,\sqrt{m_1 m_2}& \alpha\, b_{3/2}^{1}(\alpha)\,m_1 
\end{pmatrix} \label{M}
\end{equation}
and $n_{2}$ is the outer planet's mean motion $\sqrt{GM_\star/a_{2}^3}$.
We retain factors of $\alpha$ in the secular evolution to later consider the secular influence of distant perturbers on tightly spaced planets (Sec.\:\ref{sec:distantlimit}).
For cases considered below with more than two planets, we evaluate the matrix $\mathcal{M}$ numerically using the open-source \texttt{celmech} package\footnote{\url{https://github.com/shadden/celmech}.}.

The Laplace-Lagrange equations of motion in canonical variables read
\begin{equation}
\frac{d}{dt}\begin{pmatrix}\boldsymbol{G_1} \\ \boldsymbol{G_2}\end{pmatrix}
 = -i \mathcal{M} \cdot \begin{pmatrix}\boldsymbol{G_1} \\ \boldsymbol{G_2}\end{pmatrix}.
\end{equation}

Since the matrix $\mathcal{M}$ is real and symmetric, this set of linear differential equations can be solved by finding a rotated basis in which the matrix $\mathcal{D} = \mathcal{R}\cdot\mathcal{M}\cdot\mathcal{R}^T$ is diagonal. The variables rotated into this basis
\begin{equation}
\begin{pmatrix}\boldsymbol{A_1} \\ \boldsymbol{A_2}\end{pmatrix} = \mathcal{R}\cdot\begin{pmatrix}\boldsymbol{G_1} \\ \boldsymbol{G_2}\end{pmatrix} \label{Ai}
\end{equation}
are the Laplace-Lagrange modes, whose amplitudes $A_m$ remain fixed, and rotate with frequencies given by the corresponding eigenvalues of $\mathcal{D}$.
The evolution of the $\boldsymbol{G_i}$ can thus be decomposed into linear combinations of these modes, which each rotate uniformly with constant amplitude. 

Physically, the components of the matrix $\mathcal{R}_{mp}$ represent how much of mode $\boldsymbol{A_m}$ is in the planet with index $p$. 
These matrix elements are set by the planetary masses and semimajor axes, while the initial eccentricities and pericenter orientations set the conserved mode amplitudes $A_m$ as well as the initial phases (Eq.\:\ref{Ai}). 

Assuming there are no secular resonances between these frequencies, the maximum value of $G=|\boldsymbol{G}|$ for the planet with index $p$ can then be simply estimated as the sum of the contributions to the planet from each of the Laplace-Lagrange mode amplitudes, 
\begin{equation}
\Big(G_p\Big)_{\rm{max}} = \sum_{m=1}^N |\mathcal{R}_{pm}|A_m.
\end{equation}

\subsection{A Canonical Rotation} \label{canrot}
To estimate the widths of MMRs, we specifically require the eccentricity combination $\boldsymbol{\tilde{e}_{12}} \propto \boldsymbol{e_2} - \boldsymbol{e_1}$ (Eq.\:\ref{etilde}). We can study the evolution of this particular combination of eccentricities through a mass-dependent canonical rotation \citep{Wisdom86, Batygin13, Hadden19}. Defining
\begin{equation}
\begin{pmatrix} 
\boldsymbol{F_1} \\ \boldsymbol{F_2}
\end{pmatrix}
=
\mathcal{R}_\psi \cdot
\begin{pmatrix}\boldsymbol{G_1} \\ \boldsymbol{G_2}\end{pmatrix},
\label{rotation}
\end{equation}
where $\mathcal{R}$ represents a (4-dimensional) counterclockwise rotation through an angle $\psi = \pi + \sin^{-1} (\Lambda_{1}/(\Lambda_{1}+\Lambda_{2}))^{1/2}$:
\begin{equation}
\mathcal{R}_\psi
=
\begin{pmatrix}
\cos{\psi} & -\sin{\psi}\\
\sin{\psi} & \cos{\psi}
\end{pmatrix}
= -
\begin{pmatrix}
\sqrt{\frac{\Lambda_{2}}{\Lambda_{1}+\Lambda_{2}}} & -\sqrt{\frac{\Lambda_{1}}{\Lambda_{1}+\Lambda_{2}}}\\
\sqrt{\frac{\Lambda_{1}}{\Lambda_{1}+\Lambda_{2}}} & \sqrt{\frac{\Lambda_{2}}{\Lambda_{1}+\Lambda_{2}}}
\end{pmatrix} \label{Rpsi}
\end{equation}

One can check through direct substitution from Eq.\:\ref{scaling} into Eq\:\ref{rotation} that
\begin{eqnarray}
\boldsymbol{\tilde{e}_{12}} &=& \sqrt{\frac{\Lambda_1 + \Lambda_2}{\Lambda_1 \Lambda_2}} \frac{\boldsymbol{F_1}^*}{e_{12}^{\rm{cross}}} \nonumber \\
\boldsymbol{e_{com}} &\approx& \frac{\Lambda_1 \boldsymbol{e_1} + \Lambda_2 \boldsymbol{e_2}}{\Lambda_1 + \Lambda_2} = - \frac{\boldsymbol{F^*_{2}}}{\sqrt{\Lambda_1 + \Lambda_2}},
\label{F}
\end{eqnarray}
where the approximate equality in the second line holds when $\alpha \approx 1$ in the compact limit.
Thus, the new complex canonical coordinates $\boldsymbol{F}$ are scaled versions of the center-of-mass eccentricity and the $\boldsymbol{\tilde{e}_{12}}$ we require to estimate the MMR widths.

This yields a Hamiltonian in the new variables
\begin{eqnarray}
\mathcal{H'} &=& \begin{pmatrix}\boldsymbol{F_1} & \boldsymbol{F_2}\end{pmatrix} \cdot  {\mathcal{M}'}\cdot \begin{pmatrix}\boldsymbol{F_1^*} \\  \boldsymbol{F_2^*}\end{pmatrix} \\ \nonumber
\mathcal{M}' &=& \mathcal{R}_\psi \cdot{\mathcal{M}}\cdot\mathcal{R}_\psi^T; \label{Mprime}
\end{eqnarray}
we will call $\mathcal{M}'$ the rotated Laplace-Lagrange matrix.

\subsection{A Nearly Conserved Quantity for 2 Planets}

We now show that the rotation $\mathcal{R}_\psi$ from Poincar\'e variables to $\boldsymbol{F}$ is close to the matrix that diagonalizes $\mathcal{M}$.
This implies that the combinations $\boldsymbol{\tilde{e}_{12}}$ and $\boldsymbol{e_{\rm{com}}}$ are approximately the conserved secular eigenmodes of the system.

After some algebraic manipulation, using Eqs. \ref{M} and \ref{Rpsi}, the rotated Laplace-Lagrange matrix can be written as
\begin{equation}
\mathcal{M}' = - \begin{pmatrix}
\omega_{12} & 0 \\
0 & 0
\end{pmatrix}
+ d \omega_{12} \mathcal{M}_d,
\end{equation}
where, since we will show that $d$ is a small parameter, $\omega_{12}$ is approximately the eigenvalue of $\boldsymbol{F_1}$, and thus close to the precession rate of $\boldsymbol{\tilde{e}_{12}}$ (Eq\:\ref{F}),
\begin{equation}
\omega_{12} = \Bigg(\frac{\alpha\,m_1 + \alpha^{1/2}\,m_2}{M_\star}\Bigg)\frac{ b_{3/2}^1(\alpha)}{4}\,n_2. \label{w12}
\end{equation}
The parameter $d$ is the fractional difference between the two Laplace coefficients appearing in $\mathcal{M}$,
\begin{equation}
d = \frac{b_{3/2}^1(\alpha) - b_{3/2}^2(\alpha)}{b_{3/2}^1(\alpha)} \label{d}
\end{equation}
and
\begin{equation}
\mathcal{M}_d = 
\begin{pmatrix}
2\frac{\Lambda_1 \Lambda_2}{(\Lambda_{1}+\Lambda_2)^2} &
\frac{\sqrt{\Lambda_1 \Lambda_2}}{\Lambda_1 + \Lambda_2} \frac{\Lambda_1 - \Lambda_2}{\Lambda_1 + \Lambda_2} \\
\frac{\sqrt{\Lambda_1 \Lambda_2}}{\Lambda_1 + \Lambda_2} \frac{\Lambda_1 - \Lambda_2}{\Lambda_1 + \Lambda_2} & -2\frac{\Lambda_1 \Lambda_2}{(\Lambda_{1}+\Lambda_2)^2}
\end{pmatrix} \label{Mpsi}
\end{equation}
has elements that are $\mathcal{O}(1)$ or smaller.

Using an approximation introduced by \cite{Goldreich81}, we find
\begin{equation}
b_{3/2}^j(\alpha) \approx \frac{2|j|}{\pi(1-\alpha)}K_1(|j|(1-\alpha)), \label{bgt}
\end{equation}
where $K_1$ is a modified Bessel function of the second kind. 
Given that our expressions only involve $j=1$ and $j=2$, in the compact limit the argument $j(1-\alpha) \ll 1$, so we can use the asymptotic expansion near zero for $K_1(z) \approx z^{-1}$, yielding
\begin{equation}
b_{3/2}^j(\alpha) \approx \frac{2}{\pi(1-\alpha)^2}, \: j(1-\alpha) \ll 1, \label{bsimple}
\end{equation}
independent of $j$. In fact, we find Eq.\:\ref{bsimple} provides a closer approximation to $b_{3/2}^2$ than Eq.\:\ref{bgt} for $\alpha > 0.75$, and for $b_{3/2}^1$ for $\alpha > 0.33$, within 10\% of the exact value.
In this approximation, $d$ vanishes, so $d \lesssim 0.1$ for $\alpha > 0.75$, and drops to zero as $\alpha$ approaches unity (for reference $\alpha > 0.63$ inside the 2:1 MMR). 

The vanishing second eigenvalue in Eq.\:\ref{Mpsi} implies that both the magnitude and direction of $\boldsymbol{F_2} \propto \boldsymbol{e_{\rm{com}}}$ are conserved in the limit where $d$ is negligible, as we know must be the case in the compact limit from the discussion in the introduction to this section.
At finite separations, the conservation is not perfect, but the oscillations in $\boldsymbol{F_1}$ are attenuated by the factor $d < 0.1$, which suppresses variations in the widths of MMRs.

This near symmetry allowed \cite{Hadden18} to consider the resonant dynamics of two-planet systems ignoring any secular evolution.
However, additional planets break this symmetry and render the secular evolution important.

\section{Multiplanet Systems} \label{sec:multi}

Two separate effects must be considered for multiplanet systems. 
As mentioned above, one must calculate how much the secular eccentricity oscillations will modulate the stability boundary.
Second, one must account for MMRs with the additional planets in the system.
To better illustrate the dynamics, in what follows, we will first isolate each of these effects. Then, in Sec.\:\ref{sec:disc}, we will put them together.

\subsection{MMRs with Additional Planets} \label{sec:MMRs}

The approach by \cite{Hadden18} of calculating an optical depth of MMRs between a pair of planets is straightforward to generalize to the case with additional planets.
For a given planet, we now show that the simplest approximation---simply adding the MMR optical depths (Eq.\:\ref{tauH18}) with all other planets---provides a reliable estimate of the chaotic boundary.
We begin by considering the case of three planets, and will argue in Sec.\:\ref{sec:additional} that considering adjacent trios of planets in higher multiplicity systems is a useful simplification.
A new complication for three-planet systems is that the MMR optical depths of the middle planet with each of its adjacent neighbors depend on both $\tilde{e}_{12}$ and $\tilde{e}_{23}$, and these will generally evolve with time due to the secular dynamics.

We can nevertheless isolate the interaction of MMRs between different pairs of planets and visualize the results by considering a three-planet system initialized in a single Laplace-Lagrange eigenmode.
Exciting only a single eigenmode implies that, at the level of our leading-order approximation, all eccentricities, and thus $\tau_{12}$ and $\tau_{23}$, remain constant in time.

We seem to be left with four dimensions: the two dynamical separations between adjacent planets (Eq.\:\ref{S}), and $\tilde{e}_{12}$ and $\tilde{e}_{23}$ (Eq.\:\ref{etilde}).
However, the ratios between the various orbital eccentricities (and thus $\tilde{e}_{12}/\tilde{e}_{23}$) are actually fixed by the eigenmode components, and only depend on the planetary masses and orbital periods \citep[e.g.,][]{Murray99}.
If we additionally fix the initial orbital periods of the innermost and outermost planets, then choosing the middle planet's orbital period $P_2$ sets both $S_{12}$ and $S_{23}$.
The system can then specified by two parameters $(S_{12}, \tilde{e}_{12})$, just as in the two-planet problem considered by \cite{Hadden18}, and we can similarly calculate a one-dimensional optical depth, summing the contributions from each pair of adjacent planets.

\begin{figure}
    \centering \resizebox{\columnwidth}{!}{\includegraphics{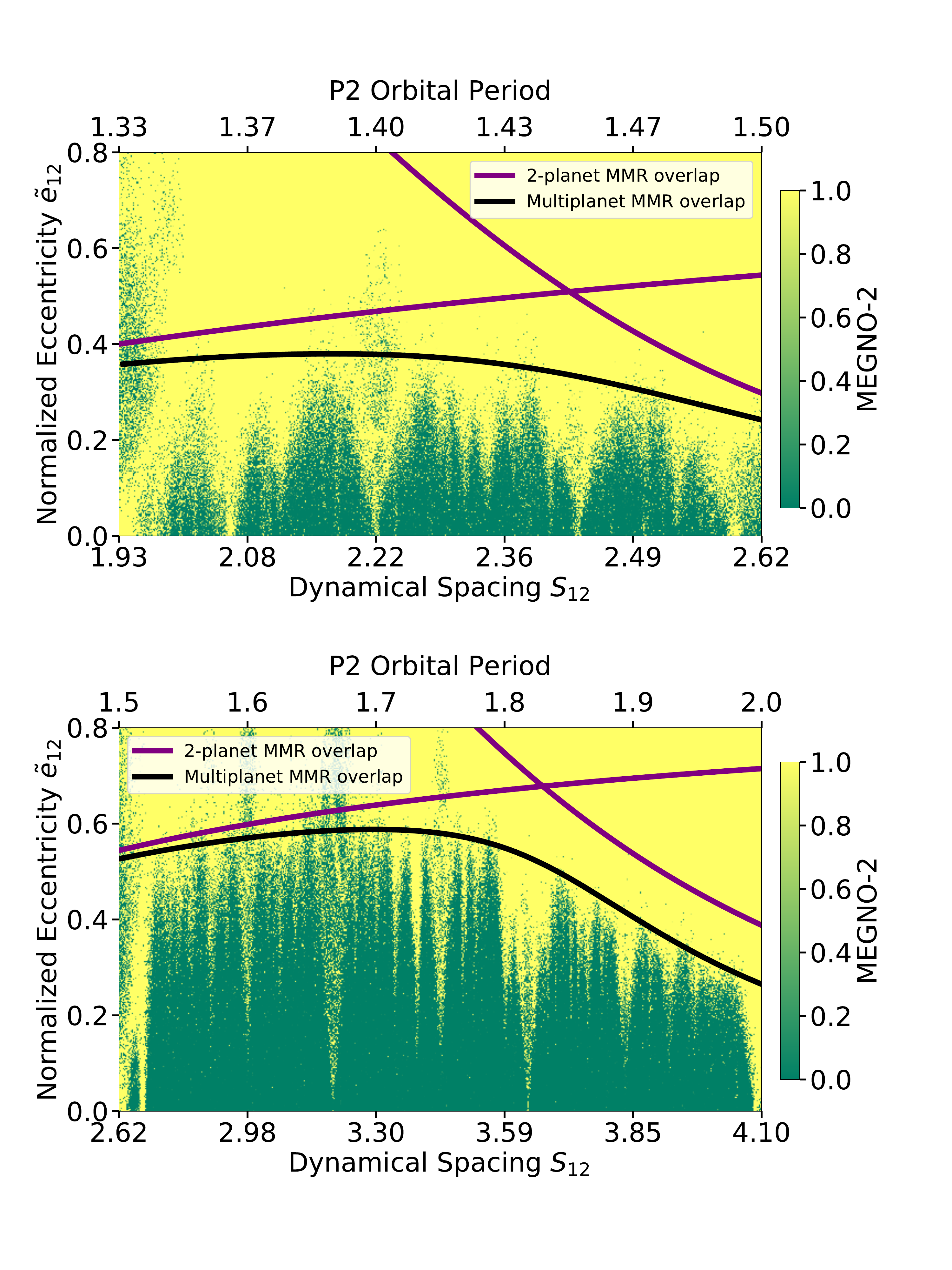}}
    \caption{The contour where the total MMR covering fraction with all pairs of planets crosses unity (black) provides an approximate boundary for chaotic motion. These integrations involved three 10\,$M_\oplus$ planets. The x-axis shows the middle planet's orbital period; the outer planets' periods were held fixed (top panel $P_1=1$, $P_3=2.033$, bottom panel $P_1=1$, $P_3=3.03$). To suppress any secular evolution and highlight the MMR optical depth criterion, the eccentricity vectors for the three planets were chosen to excite only one Laplace-Lagrange eigenmode, and to yield the value of $\tilde{e}_{12}$ on the y-axis. Purple lines denote the critical eccentricity contours at which the optical depth of MMRs between the middle planet and one of its neighbors would reach unity.
    \label{fig:3pcriterion}}
\end{figure}

In order to have both adjacent planet pairs contribute comparably to the MMR optical depth, we consider a system of three $10 M_\oplus$ planets.
We first consider a tightly spaced case, with the inner planet period fixed at $P_1=1$, and the outer period fixed at $P_3=2.033$ (with the particular value chosen to avoid a strong MMR between in the innermost and outermost planet).
We then vary the middle planet's orbital period in the range $P_2 = [1.33, 1.5]$, which approximately symmetrically covers the region between the 4:3 and 3:2 MMR with both pairs of planets.
For each simulation, we solve for the Laplace Lagrange eigenmodes at the given masses and separations.
We then initialize the system purely in the eigenmode with the fastest eigenfrequency, with the appropriate amplitude to yield the value of $e_{12}/e_{\rm{cross}}$ plotted on the y-axis in Fig.\:\ref{fig:3pcriterion}.
The eigenmode components allow us to compute the appropriate (and different) initial eccentricity vector for the third planet.

For each configuration we run a numerical integration using the \texttt{WHFast} integrator in the \texttt{REBOUND} N-body package, using a fixed timestep of $5\%$ of the pericenter passage timescale, in order to accurately integrate eccentric orbits \citep{Wisdom15}.
In each case we calculate the Laplace-Lagrange eigenfrequencies.
In compact configurations, one of the Laplace-Lagrange eigenfrequency is much slower than the other ones, reflecting the near conservation of the system's center-of-mass eccentricity.
We therefore choose to run the integrations for three times the period of the next slowest eigenfrequency of the system, which corresponds to $\sim 10^4$ orbital periods for the masses and separations we consider.
We then measure the Mean Exponential Growth of Nearby Orbits (MEGNO) chaos indicator \citep{Cincotta03}, which converges to a value of two for regular orbits and grows beyond two at a rate given by the fastest Lyapunov exponent for chaotic orbits.
Like any chaos indicator, the MEGNO can only identify chaotic trajectories with Lyapunov timescales shorter than the time of integration.
Since in all our plots we seek to study the boundary of trajectories that are chaotic on secular timescales (which we use to set the timespan of the integrations), we stretch the color map to highlight small deviations from two.
However, the results are visually indistinguishable even if we extend the color bar by an order of magnitude.
We explore this further and justify our choice of integration times in Appendix \ref{sec:megnonbody}.

The top panel of Fig.\:\ref{fig:3pcriterion} shows the resulting distribution of MEGNO values.
Yellow regions are chaotic, while dark green regions are regular.
In purple we plot contours for the two-planet optical depths from \cite{Hadden18} for each adjacent pair of planets, showing the critical eccentricity at which the MMR covering fraction reaches unity.
In black we show the contour where the sum of the two-planet optical depths crosses one.
The agreement on the left of the plot seems poor due to two wide first-order resonances (the 4:3 between the inner two planets at $P_2\approx 1.33$, and the 3:2 between the outer two planets at $P_2 \approx 1.36$), but one finds regular (green) configurations between them for eccentricities approaching the black line where the two MMRs start crossing, as expected.

In the bottom panel of Fig.\:\ref{fig:3pcriterion} we show an identical experiment but with a wider spacing. 
The inner planet remains at $P_1=1$, but the outer planet has been moved outward to $P_3=3.03$, while the middle planet now approximately symmetrically samples the range between the 3:2 and 2:1 MMR with both its neighbors.
We see that considering the total MMR optical depth for the middle planet from its MMRs with both its adjacent neighbors provides a good approximation to the chaotic boundary.

Both panels in Fig.\:\ref{fig:3pcriterion} use the general expression for the MMR optical depth (Eq.\:\ref{tauH18}) from \cite{Hadden18}.
While the low-eccentricity approximation (Eq.\:\ref{lowe}) provides nearly indistinguishable results for the three-planet criterion in the top panel, it significantly underestimates the optical depths in the lower panel where the normalized eccentricities in the regular regions extend $\gtrsim 0.4$, resulting in a boundary that is too high.

\subsection{Considering Only Adjacent Trios of Planets} \label{sec:additional}

When generalizing to even higher multiplicities, we now show that a useful further simplification is to only consider adjacent trios of planets.
Given that $\tau$ scales only weakly with mass (Eq.\:\ref{lowe}), we imagine for simplicity a system of equal-mass planets, and adopt equal spacings as the worst-case scenario for ignoring the contributions from non-adjacent planets.

Taking the eccentricities as approximately equal in such a five-planet case, the quadratic scaling of $\tau$ with $S$ (Eq.\:\ref{lowe}) implies that the middle planet would have a $\sim 25\%$ contribution to its total MMR optical depth from non-adjacent planets that are twice as far away as the adjacent planets.
In more general cases, the correction to only considering the most dynamically packed trio can be significantly smaller.
Additionally, once the period ratio with non-adjacent planets exceeds 2, one is no longer in the compact limit and Eq.\:\ref{lowe} significantly overestimates the optical depth due to the lack of first-order MMRs.
Thus, while for low-planet masses and exotic configurations one can construct counterexamples (e.g., a closely packed system with two Jupiters on the ends and two Earths in between), it is generally sufficient to consider only adjacent planets.

\begin{figure}
    \centering \resizebox{\columnwidth}{!}{\includegraphics{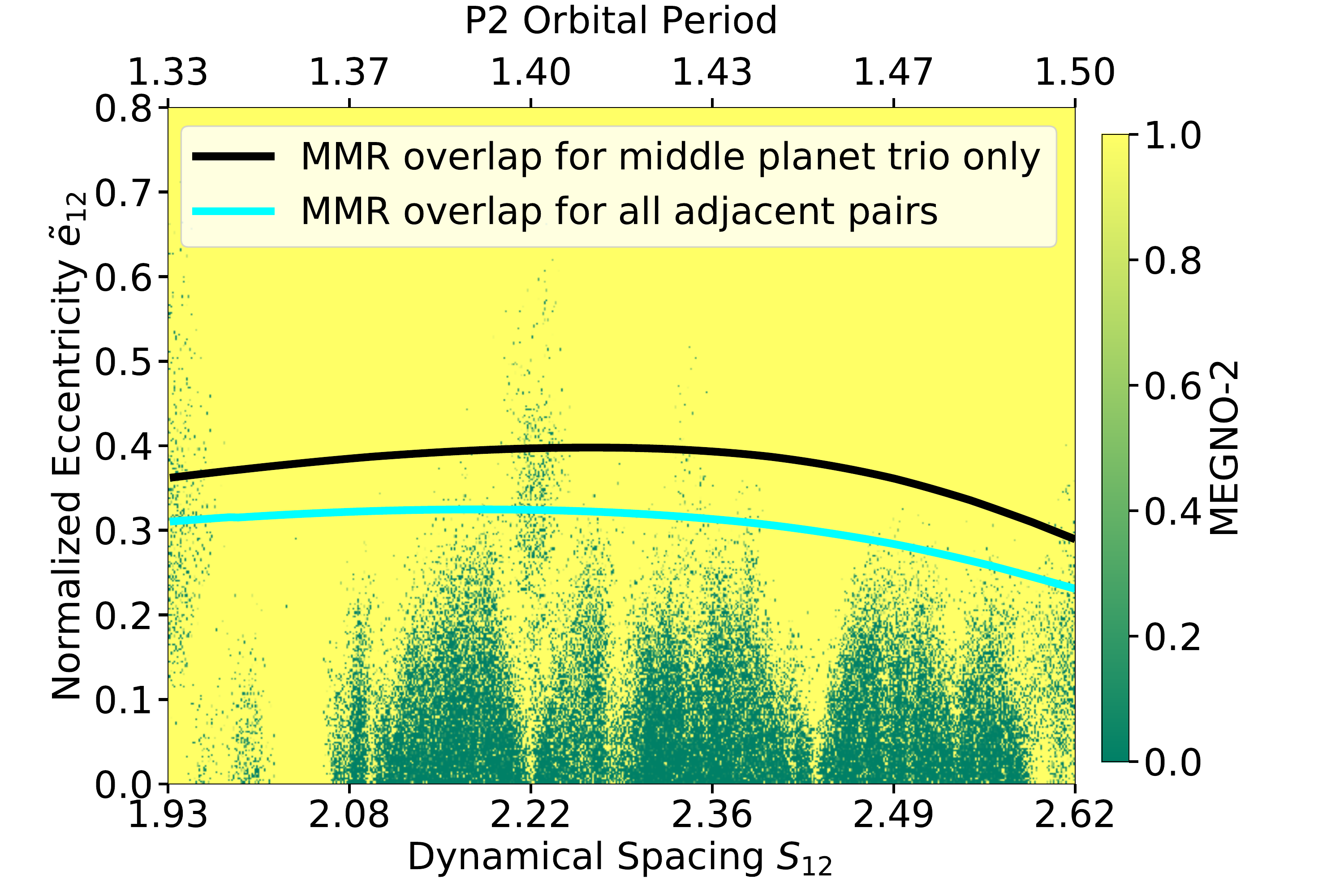}}
    \caption{Considering MMR overlap only for the tightest trio of planets approximates the chaotic boundary. Same numerical experiment as the top panel of Fig.\:\ref{fig:3pcriterion}, but with two additional planets, one interior and one exterior to the previously considered system (see text). The orbital periods are chosen as a ``worst-case scenario" such that the middle trio has approximately the same spacing as the innermost trio and outermost trio on the left and right of the plot, respectively. The eccentricity vectors are similarly initialized to only excite a single Laplace-Lagrange eigenmode and suppress the secular evolution. The MMR overlap criterion considering only the middle trio of planets (black) is within $\approx 25\%$ of the optical depth considering all adjacent pairs of planets (cyan) and the numerical results in this worst-case scenario. 
    \label{fig:5ptest}}
\end{figure}

In Fig.\:\ref{fig:5ptest}, we reproduce the numerical experiment in the top panel of Fig.\:\ref{fig:3pcriterion}, except we insert two additional $10 M_\oplus$ planets at orbital periods of $0.63$ and $3.08$. 
These values were chosen to ensure that the middle trio of planets is always the most tightly packed one (the middle trio becomes comparably spaced to the innermost and outermost trio on the left and right sides of the plot, respectively).
As above, we solve for the Laplace-Lagrange eigenmodes and initialize the system in a single mode to suppress the secular evolution.

We see that the result is qualitatively similar, with our MMR overlap criterion considering only the middle trio of planets (in black) within $\approx 25\%$ of the optical depth considering all adjacent pairs of planets (cyan) as motivated above.
The strongly chaotic region on the lower left of the top panel of Fig.\:\ref{fig:3pcriterion}, caused by the wide first-order MMRs in that period range as discussed above, is expanded in Fig.\:\ref{fig:5ptest} by the additional planets. Apart from this widening,
the chaotic structure is qualitatively similar to Fig.\:\ref{fig:3pcriterion}.
Given that we designed this numerical experiment as a worst-case scenario for the adjacent-planet approximation, we conclude that only considering the trio of adjacent planets with the highest $\tau$ provides a useful reduction in the dimensionality for high multiplicity systems.
However, we mention that even if non-adjacent planets do not contribute meaningfully to the MMR optical depth directly, they can still significantly alter the secular evolution, as we now explore.

\subsection{Secular Evolution} \label{sec:secular}

Next, we isolate the effects of secular evolution. We consider the effect of adding a third body to a system of two tightly packed planets.
We choose to place the third body at a period ratio $>2$ with the remaining planets, suppressing the optical depth of two-body MMRs it contributes due to the lack of first-order MMRs beyond the 2:1 (Sec.\:\ref{sec:MMRs}).
This allows us to focus solely on the effects of the secular evolution.

We begin by estimating the separation at which one can ignore the secular effects of unseen planets. 
This is an important question when trying to draw conclusions about the close-in systems we predominantly discover, given our difficulties in observationally detecting distant planets.

\subsubsection{When Do Distant Unseen Planets Matter?} \label{sec:distantlimit}

We found in Sec.\:\ref{canrot} that in the limit where the third planet can be ignored, $\boldsymbol{F_1}$ is approximately an eigenmode of the system.
This secular problem has been considered in a variety of contexts, most recently in understanding the excitation of mutual inclinations by an exterior perturber \citep{Becker16, Lai17}\footnote{
While we are interested in the eccentricity rather than the mutual inclination evolution, \cite{Pu18} shows that in both cases one can use the same dimensionless parameter to describe the coupling strength of the third body.}.
Rather than rederive the same results in a different way, we restate them in the context of this approach.

There are two separate regimes.
At wide separations, whenever the ratio $\epsilon$ of the {\it differential} precession induced by the distant perturber on the inner two orbits is much smaller than the mutual precession induced by the inner two orbits on one another (Eq.\:\ref{w12}), the inner two orbits precess together rigidly, i.e., at the same average rate \citep[e.g.,][]{Lai17}.
In this widely separated limit, $\boldsymbol{e_{12}}$, which tracks the differential evolution of the inner two planets' eccentricity vectors, is an eigenmode of the secular dynamics, and circulates with constant amplitude.
$\boldsymbol{F_2}$ tracks the ``rigid-body" evolution of the inner system, and will oscillate in amplitude, but has no effect on the MMR widths \citep{Hadden19}.
The widths of the MMRs between the compact pair therefore remain constant and the distant perturber has no effect on the stability boundary.

When the ratio $\epsilon$ exceeds unity, the perturber disrupts the rigid-body evolution of the inner system.
In the limit we are interested in, the inner two planets are compact ($\alpha_{12} \approx 1$) and the Laplace coefficients can be approximated through Eq.\:\ref{bsimple}.
Inserting Eq.\:\ref{bsimple} into the expression given by \cite{Lai17} and setting $\alpha_{12} \approx 1$, we have
\begin{equation}
\epsilon = \frac{9\pi}{4} \Bigg(\frac{m_p}{m_1+m_2}\Bigg)\Bigg(\frac{P_2}{P_p}\Bigg)^2 e_{12\rm{cross}}^3, \:\:\rm{(Outer}\:\rm{ Perturber)} \label{eps}
\end{equation}
where $P_2$ and $P_p$ are the periods of the middle and perturbing planet, and $e_{12\rm{cross}}$ is the crossing eccentricity of the inner two planets (Eq.\:\ref{ecross}).
We note that \cite{Denham19} derived the a similar condition under which a third body can strongly influence the secular dynamics of an inner pair of planets, but they were focused on the regime where all planets are well separated. 
In our case where the inner pair of orbits are compact (period ratio $<2$), Eq.\:\ref{eps} differs from the criterion of \cite{Denham19} by over an order of magnitude.

Finally, we mention that at short orbital periods, general relativity and other short-range forces can also suppress secular oscillations \citep[e.g.,][]{Liu15, Wei21, Faridani21}.
For closely packed systems, it is again important to consider whether the {\it differential} precession on the two orbits exceeds their mutual gravitationally induced precession.

\subsubsection{Secular Modulation of the Stability Boundary} \label{sec:secularmod}

\begin{figure}
    \centering \resizebox{\columnwidth}{!}{\includegraphics{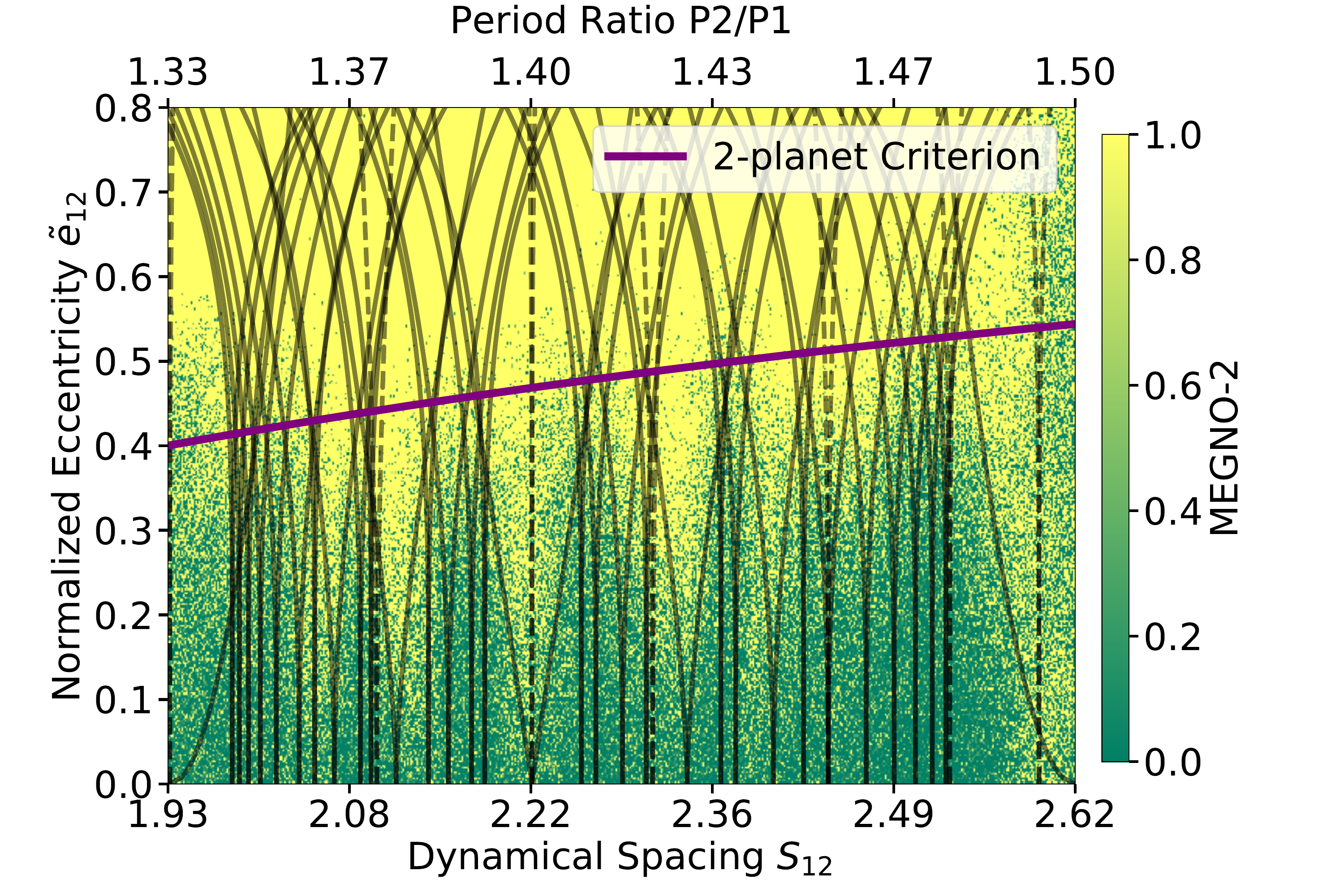}}
    \caption{Control experiment of a three-planet system that follows the 2-planet chaotic boundary. Two 10 $M_\oplus$ planets with inner period P1 fixed at 1, and P2 varied along the x-axis. There is a 1 $M_{\rm{Jup}}$ perturber always at a period P3=3.2. The eccentricities and pericenters are chosen to excite only one Laplace-Lagrange mode, suppressing the secular evolution. The MMRs between the middle and outer planet (shown in dashed lines) do not meaningfully contribute to the total optical depth of MMRs. This three-planet system therefore still exhibits the two-planet stability boundary (purple) from \cite{Hadden18} corresponding to the Jupiter-mass planet not being present.
    \label{fig:singleeigen}}
\end{figure}

The MMR widths between a compact pair of planets, and thus the two-planet stability criterion of \cite{Hadden18}, depend only on the eccentricity combination $\tilde{e}_{12}$ (Eq.\:\ref{etilde}).
As the MMR widths expand and contract through the secular eccentricity evolution induced by a distant third planet, the task is therefore to determine the maximum value $e_{12}^{\rm{max}}$ in the secular cycle.
it is thus useful to analyze the problem in the same rotated basis as in Sec.\:\ref{canrot},
\begin{equation}
\begin{pmatrix} \boldsymbol{F_1} \\ \boldsymbol{F_2} \\ \boldsymbol{F_3} \end{pmatrix}
=
\begin{pmatrix}
\cos{\psi} & -\sin{\psi} & 0\\
\sin{\psi} & \cos{\psi} & 0 \\
0 & 0 & 1
\end{pmatrix}
\begin{pmatrix}\boldsymbol{G_1} \\ \boldsymbol{G_2} \\ \boldsymbol{G_3}\end{pmatrix}, \label{Rpsibig}
\end{equation}
where $\boldsymbol{F_1} \propto \boldsymbol{\tilde{e}_{12}}$, $\boldsymbol{F_2} \propto \boldsymbol{e_{\rm{com}}}$ (Eq.\:\ref{F}), and $\boldsymbol{F_3} = \boldsymbol{G_3}$.

\begin{figure*}
     \resizebox{0.99\textwidth}{!}{\includegraphics{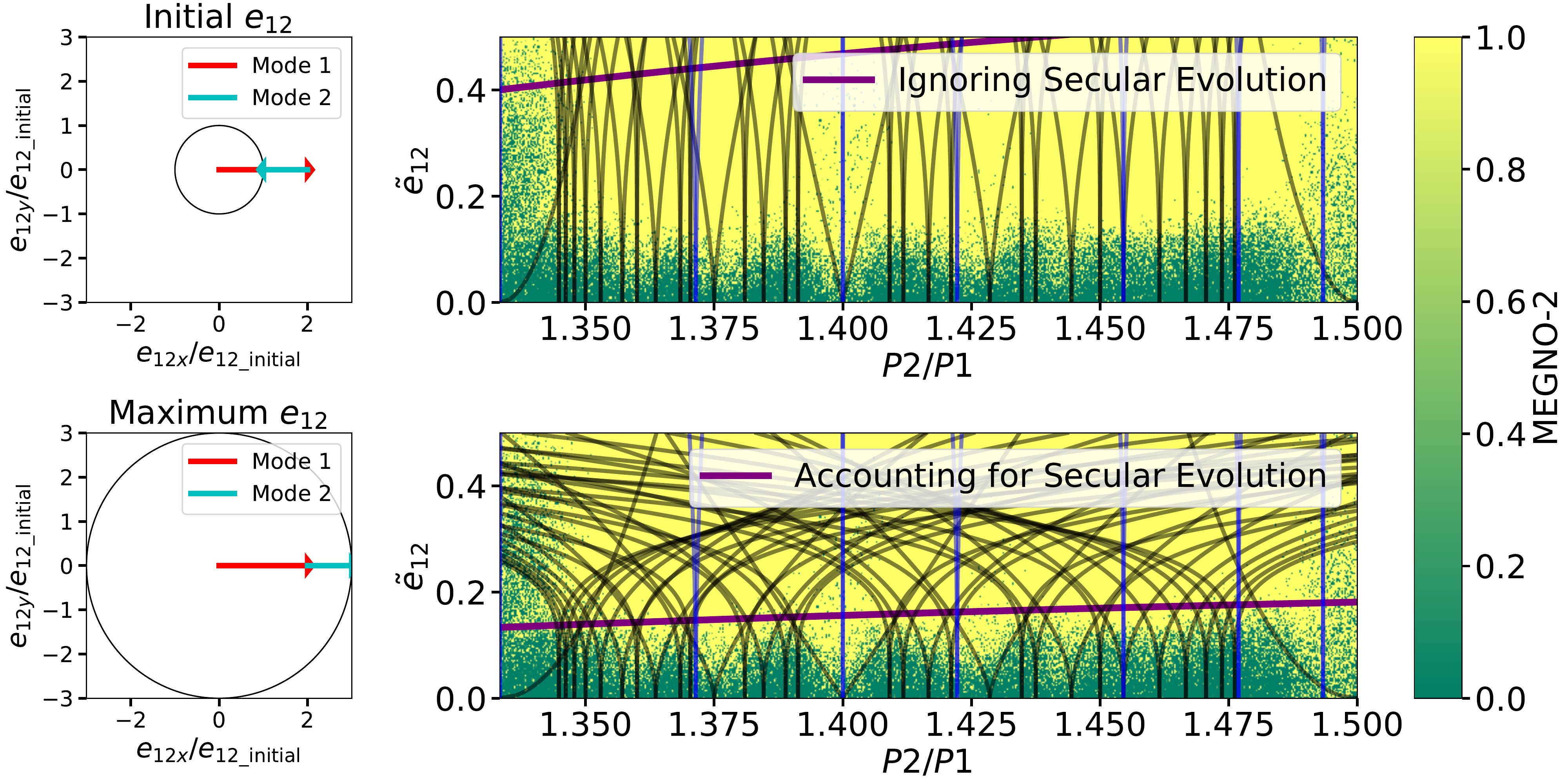}}
    \caption{Secular evolution can strongly modulate the chaotic boundary. Same planet masses and separations as in Fig.\:\ref{fig:singleeigen}, except now the eccentricities and pericenters are initialized so that $\boldsymbol{e_{12}}$ is made up of two eigenmodes with initial conditions depicted in the top left panel. The radius of the black circle given by the vector sum of the two mode contributions is $e_{12}$, and the units on the axes are the factors by which $e_{12}$ changes from its original value. While the initial $e_{12}$ changes on the y-axis of the right panels, the mode amplitudes are chosen so that $e_{12}$ always changes by the same factor of three. The top right panel shows that estimating MMR overlap using the initial eccentricities does not match the chaotic boundary. The bottom left panel shows the maximum in the eccentricity cycle, and the corresponding bottom right panel shows that the MMR widths at the secular maximum do provide an accurate overlap boundary.
    \label{fig:mixedeigen}}
\end{figure*}

We consider a suite of configurations with two tightly packed $10 M_\oplus$ planets around a solar-mass star.
The first planet is always initialized at an orbital period $P_1 = 1$, while we vary $P_2$ from the 4:3 to the 3:2 MMR with the inner planet.
In order to avoid considering MMRs with the perturbing third planet, in all configurations we place the perturber beyond the 2:1 resonance, at a period $P_3=3.2$.
In order for the perturber to drive differential precession in the inner pair of planets, we assign it a Jupiter mass.
This corresponds to $\epsilon \approx 0.1-0.3$ (Eq.\:\ref{eps}) in the period range we consider.
In other words, the differential precession induced by the perturber mixes $\approx 10-30\%$ of the center-of-mass degree of freedom $\boldsymbol{F_{2}}$ into the eigenmode that is almost purely made up of $\boldsymbol{e_{12}}$ in the absence of the perturber (Sec.\:\ref{sec:2p}).
This is enough to drive substantial modulation of the MMR widths depending on initial conditions, as we show below.

As a first control experiment, we suppress the secular evolution by initializing the eccentricities in a single eigenmode as in Sec.\:\ref{sec:MMRs}, specifically in the eigenmode associated with $e_{12\rm{com}}$.
We vary $e_{12}$ along the y-axis of Fig.\:\ref{fig:singleeigen}, while the eigenmode components set the ratio of $e_{12\rm{com}}$ and $e_3$ to $e_{12}$.
Over our range of periods for the middle planet, $e_{12\rm{com}}/e_{12}$ varies from $\approx 0.8-2.25$, while planet three remains nearly circular, with $e_3/e_{12}$ varying from $\approx 0.02-0.07$.
Thus, the MMR widths should stay fixed in time, and the prediction is that the stability boundary should follow the two-planet stability boundary of \cite{Hadden18}.

We plot the results in Fig.\:\ref{fig:singleeigen}.
In solid black lines, we plot the widths of all MMRs up to tenth order between the compact pair of planets, with widths given by $\frac{\delta P}{P} \approx \frac{3}{2} \frac{\delta a}{a}$, and the latter given by Eq.\:\ref{MMRwidth}.
In dashed black lines, we plot the widths of MMRs between the middle and outer planet, which we see have negligible widths.

There are few regular trajectories beyond the two-planet criterion in purple from \cite{Hadden18}.
The most obvious exception is inside strong MMRs, like the 4:3 on the leftmost side of the plot.
We reiterate that our approximations do not cover such cases---certainly the dynamics in those regions will not be well described by the Laplace-Lagrange secular dynamics we described above.
Many authors have considered the secular dynamics inside MMRs \citep{Wisdom85, Morbidelli93, Batygin13, Pichierri17, Hadden19}, but we do not pursue this further.

By engineering a configuration in which MMRs with the third planet can be ignored, and by initializing the eccentricity vectors so as to suppress the secular evolution, we set up a three-planet system that follows the two-planet stability boundary despite the large mass and relative proximity of the perturbing planet.

\begin{figure*}
    \resizebox{0.99\textwidth}{!}{\includegraphics{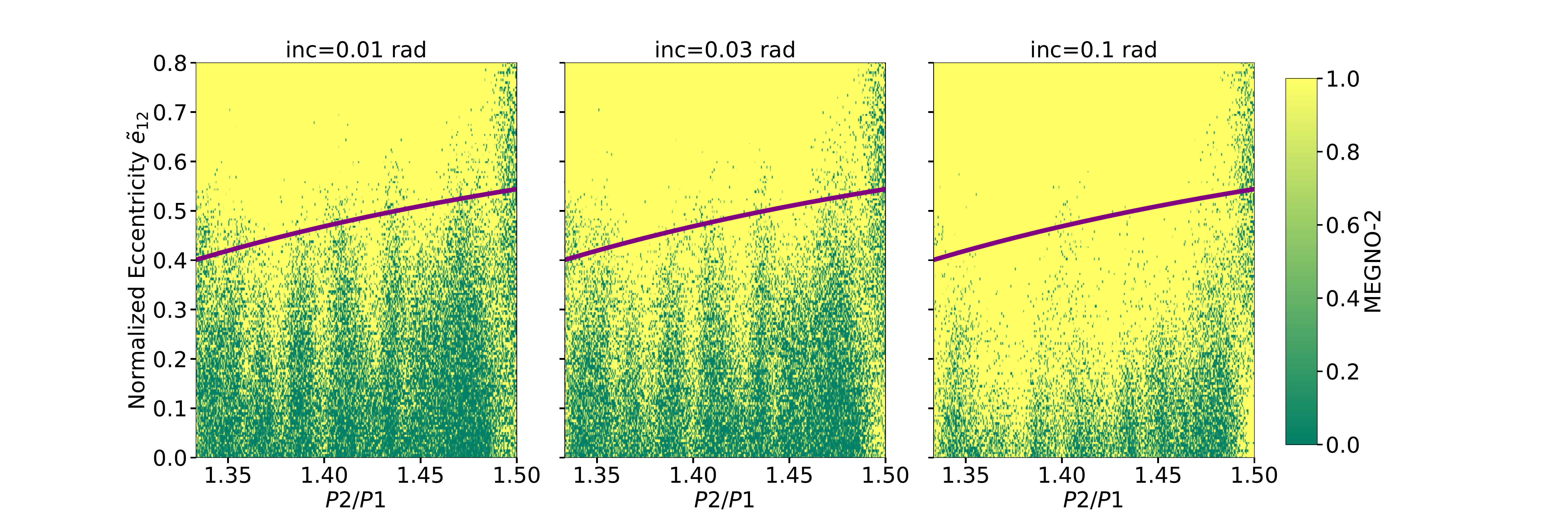}}
    \caption{Mutual inclinations within a few degrees do not modify the chaotic boundary. Same numerical experiment as in Fig.\:\ref{fig:singleeigen}, but with all planets assigned the same orbital inclination (given above each subplot), and a random longitude of ascending node. The panels thus correspond to maximum mutual inclinations of $1.1^\circ, 3.4^\circ$ and $11^\circ$, respectively.
    \label{fig:inclinations}}
\end{figure*}

In our second experiment we adopt the same masses, separations and initial $e_{12}$ as above, but this time we initialize the $e_{12}$ as a mix of the two eigenmodes associated with the compact pair of planets.
In particular, for each configuration, we first solve for the rotated Laplace-Lagrange eigenmodes $\boldsymbol{A_m}$.
For a given value of $e_{12}$ we plot on the y axis of Fig.\:\ref{fig:mixedeigen}, we then choose the mode amplitudes such that $\boldsymbol{A_1}$ contributes $2e_{12}$ to the inner pair's relative eccentricity, and $\boldsymbol{A_2}$ contributes $e_{12}$.
We then start the modes anti-aligned so the total eccentricity is $2e_{12} - e_{12} = e_{12}$ as desired (top left panel of Fig.\:\ref{fig:mixedeigen}).
We see on the top right panel that the MMR widths are too low at the initial value of $e_{12}$ to explain the chaotic regions through their overlap.

However, once the system starts evolving, and the eigenmodes rotate at their own eigenfrequencies, $e_{12}$ and thus the resonance widths will change with time. 
Given our setup, all configurations will reach a value of $e_{12}$ that is three times higher than the initial value at the maximum in the secular cycle when the modes align (bottom left panel).
In the bottom right panel we plot the MMR widths using this inflation factor of three for the eccentricities.
We also plot the boundary from \cite{Hadden18} for the critical eccentricity depressed by the same factor of three.

In this picture, the MMR widths are adiabatically modulated by the secular evolution, and the MMR criterion of \cite{Hadden18} needs to be modified so that it does not refer to the critical {\it initial} eccentricity, but rather the maximum eccentricity attained during a secular cycle.

\subsection{Ignoring Inclinations} \label{sec:inclinations}

Throughout the paper we have worked assuming co-planar orbits.
We now examine the role of small inclinations in modifying the chaotic boundary.
We reproduce the numerical experiment in Fig.\:\ref{fig:singleeigen}, with all planets starting with the same orbital inclination and randomly assigned longitudes of ascending node.
The results are shown in Fig.\:\ref{fig:inclinations}, with inclinations of $0.01, 0.03$ and $0.1$ radians as labeled above each subplot.
Given the randomly assigned node longitudes, these correspond to maximum mutual inclinations between pairs of planets of $1.1^\circ, 3.4^\circ$ and $11^\circ$, respectively.
We see that inclinations within a few degrees, typical of transiting multiplanet systems \citep[e.g.,][]{Fabrycky14} do not affect the chaotic boundary, although mutual inclinations $\gtrsim 5^\circ$ would need to be taken into account.

\section{Discussion} \label{sec:disc}

We now put everything together. We run a numerical experiment corresponding to the right panel in Fig.\:\ref{fig:schematic}.
We consider three $10 M_\oplus$ planets on equally spaced orbits and equal initial eccentricities (varied along the y-axis).
The initial orbital pericenters are selected at $120^\circ$ intervals so as to drive secular evolution by exciting all the secular eigenmodes.

\begin{figure*}
    \resizebox{0.99\textwidth}{!}{\includegraphics{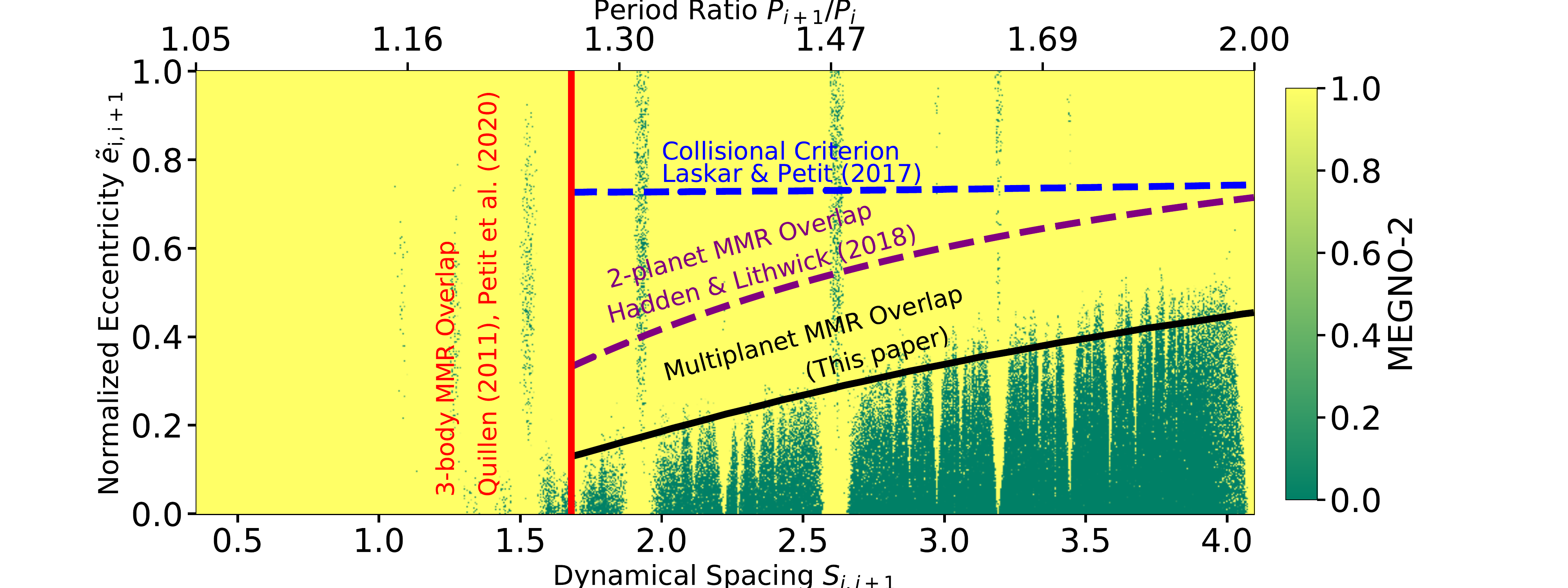}}
    \caption{The chaotic boundary in multiplanet systems. Shown are results from simulations of three equally spaced $10 M_\oplus$ planets, as a function of the dynamical spacing (Eq.\:\ref{S}) and the normalized eccentricities (Eq.\:\ref{etilde}); see the text for details. At the tightest separations (left of solid red line), chaos is driven by the overlap of 3-body MMRs \citep{Quillen11, Petit20}. At wider separations, the chaotic boundary is set by the overlap of 2-body MMRs (solid black line). We compare this boundary to the limit for two-body MMR overlap in 2-planet systems from \cite{Hadden18}, shown in dashed purple. The difference between the black and purple lines is due to the secular evolution of the eccentricities driven by the third planet. We also show in dashed blue the critical eccentricity at which the Laplace-Lagrange secular evolution would achieve the orbit-crossing condition of \cite{Laskar17}.
    \label{fig:summary}}
\end{figure*}

The result is shown in Fig.\:\ref{fig:summary}.
First, we see that in this equally spaced, equal-mass case, dynamical spacings $\lesssim 1.68$ are chaotic at all eccentricities due to the overlap of 3-body MMRs as predicted by \cite{Quillen11}.
As shown by \cite{Petit20} and various numerical experiments \citep[e.g.,][]{Faber07, Smith09, Obertas17}, the resulting instabilities occur within $\sim 10^{10}$ orbits, which is significantly shorter than most exoplanet hosts' main-sequence lifetimes.
In cases with unequal masses and/or period ratios, where adjacent planets have different dynamical spacings, this boundary for 3-body MMR overlap would shift toward smaller dynamical spacing of the tightest planet pair $<1.68$, with general expressions given by \cite{Petit20}.

We have argued that beyond the 3-body MMR overlap limit, chaos on secular timescales is driven by the overlap of 2-body MMRs, and that this boundary can be significantly modulated by the system's regular (non-chaotic) secular evolution.
The purple dashed line shows the two-planet limit for the overlap of 2-body MMRs from \cite{Hadden18}.
Our boundary for this case is in black, with approximately half the deviation from \cite{Hadden18} arising from considering MMRs with both of the middle planet's adjacent neighbors (Sec.\:\ref{sec:MMRs}), and the other half coming from the secular evolution (Sec.\:\ref{sec:secular}).
In this case, our eccentricity boundary differs from the two-planet criterion of \cite{Hadden18} by a factor of $\approx 2$, though this difference can be larger or smaller depending on the system's secular mode amplitudes (Sec.\:\ref{sec:secularmod}).

We can see that while our boundary matches the numerical results well at tight separations, it is $\approx 10\%$ too low at period ratios between the 3:2 and 2:1 resonances. 
This arises at least in part from the breakdown of the compact approximation at these wide separations in the 2-planet MMR overlap criterion (Appendix \:\ref{sec:critcomp}).
We also comment that it is not clear that our approximation of simply summing MMR widths from both adjacent pairs of planets (Sec.\:\ref{sec:MMRs}) should be valid in this equally spaced case.
In this configuration, the MMRs of the middle planet with its inner neighbor lie on top of those with its outer neighbor, so one might expect summing their widths to be double-counting their contribution to the MMR covering fraction.
While this special case might benefit from a more careful analysis, it is reassuring from Fig.\:\ref{fig:summary} that our simple approach still gives a close approximation to the chaotic boundary.
We have also checked a nearly equally spaced case where we space the outer pair 5\% wider than the inner pair to offset the MMRs between adjacent planets, and find comparable agreement to numerical results as in Fig.\:\ref{fig:summary}.

Finally, while \cite{Laskar17} and \cite{Petit17} considered very long timescales on which the secular eigenmodes might evolve through secular chaos, it is instructive to compare what they would have predicted on shorter timescales under the regular Laplace-Lagrange evolution we consider.
In particular, \cite{Laskar17} study conditions that would lead to orbit-crossing, which in the compact limit reduces to $\tilde{e} = 1$.
In dashed blue, we therefore plot the critical initial eccentricity that reaches a maximum $\tilde{e}$ of unity under our assumption of Laplace-Lagrange evolution.
We see that in this case, this estimate is too high by a factor of $\approx 2-5$ compared to the numerical results.

\cite{Petit17} also consider the possibility that the secular evolution raises the eccentricities enough for first-order MMRs to overlap. 
In this case, at separations beyond the 3-body MMR overlap region, we find that the orbits always become orbit-crossing before the eccentricities get high enough for first-order MMRs to overlap.
This is not true in general \citep{Petit17}, but the condition for overlap of the set of all two-body MMRs (dashed purple line in Fig.\:\ref{fig:summary}) always occurs at lower eccentricities than the criterion for the overlap of only first-order MMRs. 

\section{Conclusion} \label{sec:conc}

We have investigated the boundary in parameter space for chaotic motion in compact multiplanet systems on secular timescales (longer than the orbital period by roughly the star-planet mass ratio).
At the tightest orbital separations (left of the red line in Fig.\:\ref{fig:summary}), the overlap of three-body MMRs causes chaos even for initially circular orbits \citep{Quillen11, Petit20}.
At wider separations, we find that the chaotic boundary is instead set by the overlap of 2-body MMRs at finite eccentricities.
In this picture, a better measure of how tightly the system is packed than the typically used Hill separation is the dynamical spacing $S$ (Eq.\:\ref{S}) \citep{Quillen11, Hadden18, Petit20}.

We argued that a central reason why compact systems with three or more planets are chaotic over a significantly broader range of parameter space than two-planet configurations is a near-symmetry that suppresses changes in the widths of MMRs in two-planet systems (Sec.\:\ref{sec:2p}).
By contrast, configurations of three or more planets undergo long-term secular eccentricity oscillations that cause MMR widths to expand and contract, adiabatically modulating the chaotic boundary (Sec.\:\ref{sec:secularmod}).
We find that modeling this secular evolution to lowest order in the eccentricities through classical Laplace-Lagrange theory is adequate, and derive semi-analytic expressions for MMR overlap at the maximum in the secular cycle that closely match the envelopes for chaotic motion measured through N-body integrations.

A challenge for understanding the long-term stability of multiplanet systems is the high dimensionality of the problem. 
We find that two approximations significantly simplify the problem.
First, we argue in Sec.\:\ref{sec:additional} that the widths of two-body MMRs decrease fast enough with orbital separation that it is sufficient to consider adjacent trios of planets separately, and adopt the most stringent condition among all trios.
Second, we find that the coplanar problem is a good approximation for mutual inclinations below a few degrees, typical of transiting multiplanet systems (Sec.\:\ref{sec:inclinations}).

For a given trio of planets, this leaves four dimensionless parameters determining the stability of the system: the two dynamical spacings $S_{12}$ and $S_{23}$ (Eq.\:\ref{S}), and the maximum normalized eccentricities (Eq.\:\ref{etilde}) $\tilde{e}_{12}^{max}$ and $\tilde{e}_{23}^{max}$ attained during a secular cycle. 
To visualize the results in the figures in the paper we reduce this dimensionality by variously restricting the masses, spacings and orbital periods, as well as by fixing the initialization of the secular eigenmodes.
For general cases, we provide routines for calculating the chaotic boundary as part of the open-source SPOCK package \citep{Tamayo20} for community use.

We have checked that, with the exception of a minority of configurations inside strong MMRs, systems beyond our chaotic boundary are short-lived.
On timescales much longer than the secular timescale, however, it is possible for overlap of secular resonances \citep[e.g.,][]{Lithwick11secularchaos} to drive secular chaos that mixes the secular mode amplitudes and pushes the instability boundary to lower eccentricities than our short-term chaotic boundary (black line in Fig.\:\ref{fig:summary}).
This question was pursued by \cite{Laskar17} and \cite{Petit17}, who analyzed the worst-case scenario where secular chaos drives all the system's angular momentum deficit to a single pair of planets.
However, we have seen that the instability boundary they considered for the short-term dynamics (blue line in Fig.\:\ref{fig:summary}) was not accurate; their corresponding AMD criterion also requires modification. 
We derive an updated AMD criterion in Appendix \ref{sec:amd}, which shows that the AMD threshold below which stability is guaranteed by \cite{Laskar17} is lowered by a multiplicative factor that is exponential in the dynamical spacing.
It represents a significant correction for compact systems with $S \lesssim 3$.

An issue we have not resolved is the fate of planetary configurations on Gyr timescales. Is secular chaos important in driving long-term instabilities in observed compact multiplanet systems? Or does MMR overlap typically limit the eccentricities to such small values on short timescales that secular resonances are too narrow to overlap, and secular chaos is rare?
The answer to this question has important implications for the timing at which the final architectures of compact exoplanet systems are set. We plan to address these long-term dynamics in future work.

We make the scripts used to run the above numerical experiments and generate their corresponding figures available at \url{https://github.com/dtamayo/chaospaper}.
We also provide routines for evaluating our expressions and estimating whether input planetary configurations are dynamically chaotic at \url{https://github.com/dtamayo/spock} as part of the open-source SPOCK package \citep{Tamayo20}. 

\section{Acknowledgements}

We thank the referee, Antoine Petit, for a careful and insightful review.
We are also grateful to Sam Hadden and Alysa Obertas for illuminating discussions.
Support for DT was provided by NASA through the NASA Hubble Fellowship grant HST-HF2-51423.001-A awarded  by  the  Space  Telescope  Science  Institute,  which  is  operated  by  the  Association  of  Universities  for  Research  in  Astronomy,  Inc.,  for  NASA,  under  contract  NAS5-26555.
This work was supported in part by the Natural Sciences and Engineering 
Research Council of Canada (NSERC), funding reference number RGPIN-2020-03885.
This research was made possible by the open-source projects \texttt{REBOUND} \citep{Rein12}, \texttt{celmech}\footnote{\url{https://github.com/shadden/celmech}.}, \texttt{Jupyter} \citep{jupyter}, \texttt{iPython} \citep{ipython}, \citep{numpy},
and \texttt{matplotlib} \citep{matplotlib, matplotlib2}.

\appendix
\section{Chaos vs Instability Times} \label{sec:megnonbody}

The MEGNO classifies whether orbits are chaotic over the timespan of an N-body integration.
In particular, it does not preclude chaotic behavior on longer timescales.
When it identifies a configuration as chaotic, there is also no guarantee that the planets will undergo an instability within a reasonable amount of time.
Nevertheless, the MEGNO is a useful tool for rapidly surveying parameter space, and for highlighting boundaries where the dynamical behavior changes.

In Fig.\:\ref{fig:singleeigen}, we chose to run for three secular cycles of the second-slowest eigenmode, corresponding approximately to the precession period of the compact planet pair under the perturbations from the exterior Jupiter, since it is on this timescale that the MMRs evolve to their maximal widths.
If instead we run the for three secular cycles of the slowest eigenmode, corresponding approximately to the precession period of the Jupiter perturber, we find that the vast majority of the parameter space (even at low eccentricities) is chaotic.
This is true even when removing the Jupiter perturber, and integrating for three secular cycles of the slowest eigenmode for the compact pair of planets (which becomes long due to the near-symmetry in the two-planet case discussed in Sec.\:\ref{sec:2p}).
While we have not identified the source of this chaos, it is not due to overlap of separate 2-body MMRs, and it is not clear whether it can drive instabilities over long timescales, which is ultimately what we are interested in.

\begin{figure}
     \centering \resizebox{\columnwidth}{!}{\includegraphics{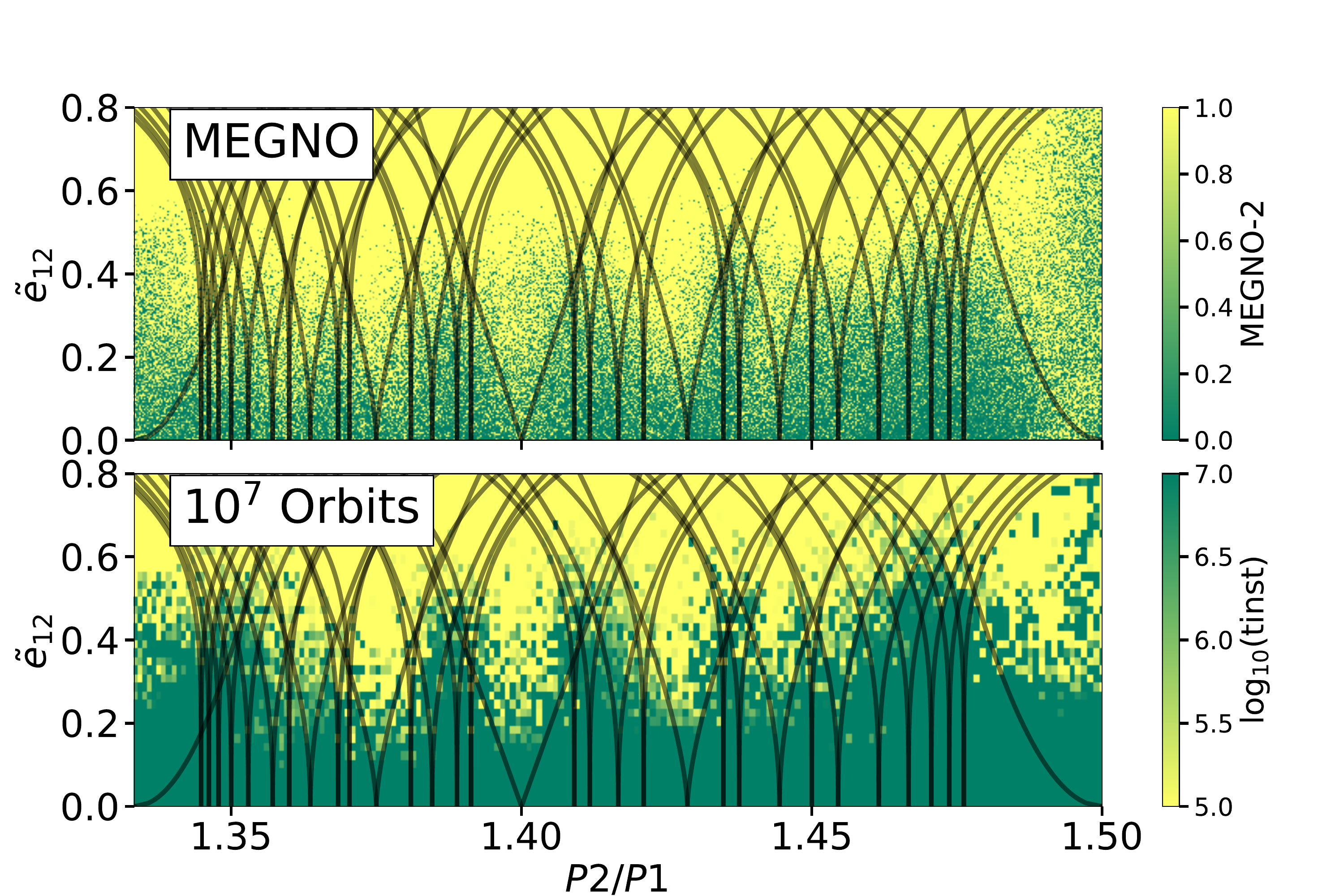}}
    \caption{ Comparison of the MEGNO map in Fig.\:\ref{fig:singleeigen} to instability times measured through integrations lasting up to $10^7$ orbits. Measuring the MEGNO over integrations spanning the secular timescales on which the MMRs expand and contract capture the appropriate chaotic boundary corresponding to short-term instabilities.
    \label{fig:megnonbody}}
\end{figure}

In Fig.\:\ref{fig:megnonbody} we measure the instability times in integrations up to $10^7$ innermost planet orbits (defined as a close encounter within the sum of planets' individual Hill radii), showing that the regions below which 2-body MMRs overlap are stable on these timescales, despite being chaotic on much shorter timescales.
Conversely, the vast majority of configurations above this boundary go unstable.
While it is unsatisfying to not have a more rigorous delineation, we conclude that measuring chaos on timescales comparable to the secular timescale on which the MMRs expand on contract secularly appropriately captures the boundary for short-term instabilities, and defer the behavior over much longer timescales to future work.
We point out that this issue also exists for previous criteria for the onset of chaos \citep[e.g.,][]{Wisdom80, Hadden18}.

\section{Eccentric Two-Planet Chaos Criterion} \label{sec:critcomp}

The value of the expressions for the MMR optical depth (Eq.\:\ref{tauH18}) and critical eccentricity for chaos (Eq.\:\ref{H18crit}) derived by \cite{Hadden18} is that they work over the full range of eccentricities.
However, given that eccentricities in compact multiplanet systems measured through TTVs are typically small \citep[e.g.,][]{Lithwick12, Jontof14, Hadden17}, it can also be useful to find simpler approximations valid at low eccentricities.

\begin{figure}
     \centering \resizebox{\columnwidth}{!}{\includegraphics{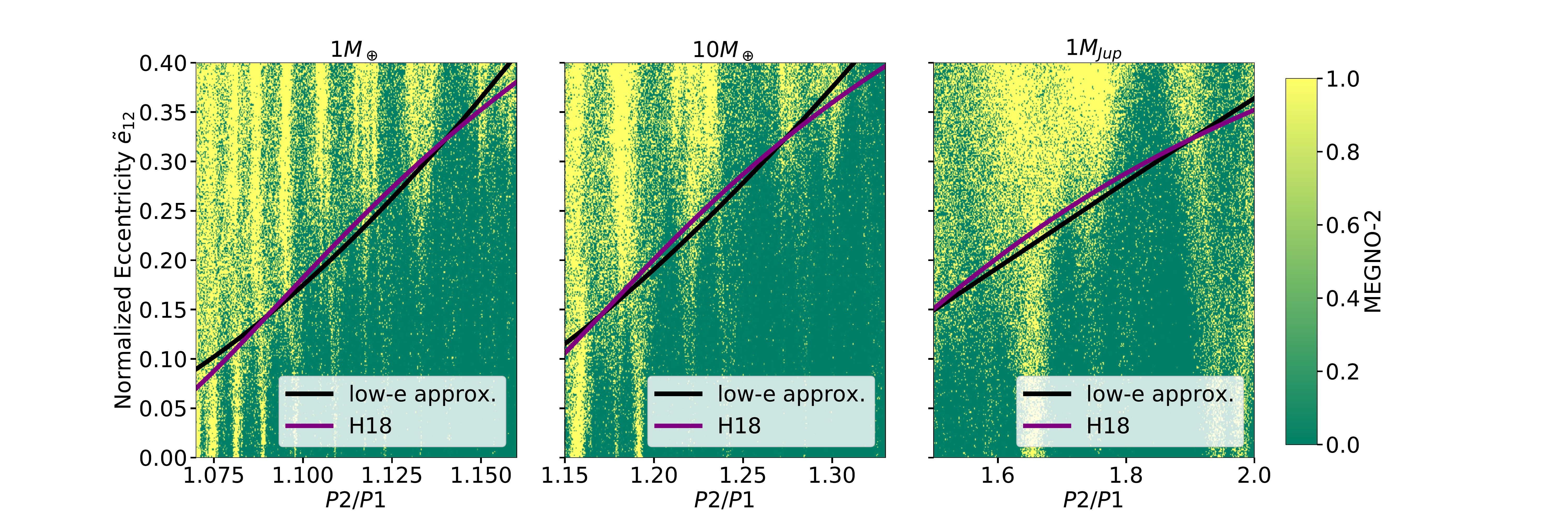}}
    \caption{ Comparison of the critical eccentricity at which 2-body MMRs overlap for two-planet systems from \cite{Hadden18} (Eq.\:\ref{H18crit}, purple) to our low-eccentricity approximation (Eq.\:\ref{lowe}, black). Both planets have the same mass (given above each subplot). The low approximation agrees with \cite{Hadden18} within 10\% for $0.1 < \tilde{e}_{12} \lesssim 0.4$.
    \label{fig:critcomp}}
\end{figure}

In Fig.\:\ref{fig:critcomp} we compare our low-eccentricity approximations to the general expressions of \cite{Hadden18} for the case of two $1 M_\oplus$ planets (left panel), two $10 M_\oplus$ planets (middle panel) and two $1 M_{\rm{Jup}}$ planets (right panel).
The period ratio ranges from the critical separation (Eq.\:\ref{wisdom}) at which first-order MMRs overlap at zero eccentricity \citep{Wisdom80}, out to approximately the period ratio where the critical eccentricity for MMR overlap (Eq.\:\ref{lowe}) reaches $\tilde{e}_{12} = 0.4$.
Integrations were run and the MEGNO was measured as described in the main text, and analogous to our other numerical experiments, we integrate for three periods of the fast Laplace-Lagrange eigenfrequency (since the nearly conserved eigenmode has a much slower eigenfrequency, Sec.\:\ref{sec:2p}).
We initialize the eccentricities purely in the fast eigenmode as in Sec.\:\ref{sec:MMRs}.
This step is not necessary at close orbital separations due to the near-symmetry discussed in Sec.\:\ref{sec:2p}, but makes a noticeable difference at wider separations between the 3:2 and 2:1 MMRs.

We find that the approximations in Eq.\:\ref{lowe} fall within $10\%$ of those from \cite{Hadden18} in the range $0.1<\tilde{e}_{12}<0.4$. 
We note that the regime at lower eccentricities is not physically relevant, since first-order MMRs typically overlap anyway below $\tilde{e}_{12} = 0.1$ (see Fig.\:\ref{fig:critcomp}, where the left-most period ratio has been chosen to match this criterion from \citealt{Wisdom80}).

\begin{figure}
     \centering \resizebox{\columnwidth}{!}{\includegraphics{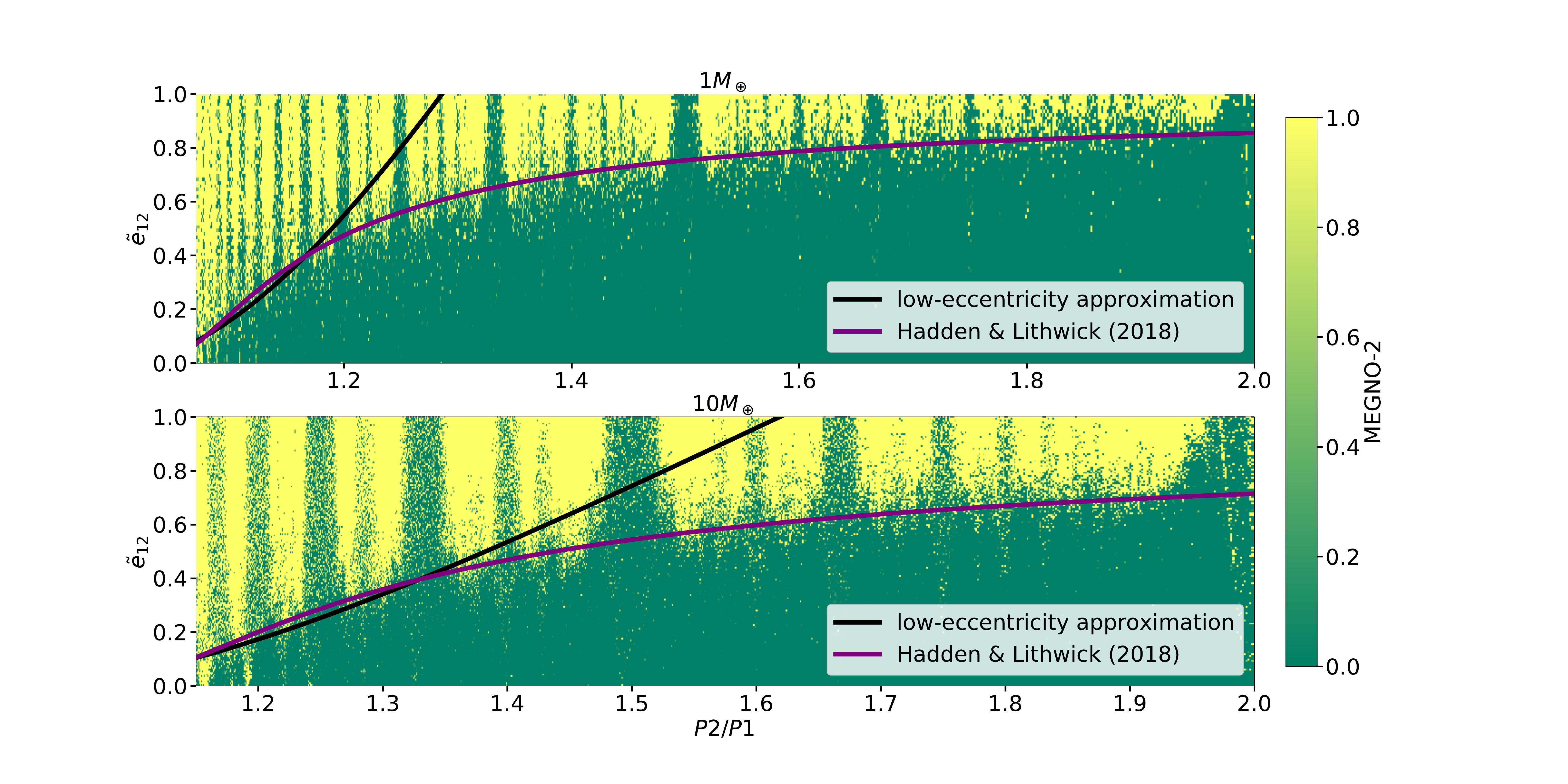}}
    \caption{ Extension of the left two panels in Fig.\:\ref{fig:critcomp} to higher eccentricities and wider period ratios, showing the failure of the low-eccentricity approximation beyond $\tilde{e}_{12} \approx 0.4$.
    \label{fig:critcompwide}}
\end{figure}

Figure \ref{fig:critcompwide} shows an extension of the left two panels in Fig.\:\ref{fig:critcomp} to higher eccentricities and wider period ratios. 
The criterion of \cite{Hadden18} matches the numerical results at all eccentricities, while the simpler low-eccentricity approximation fails beyond $\tilde{e}_{12} \gtrsim 0.4$.

\section{An Updated AMD criterion} \label{sec:amd}

It is straightforward to generalize the AMD criterion of \cite{Petit17} to calculate the minimum AMD that allows for the overlap of higher order MMRs.
This was also explored by \cite{PetitThesis}, but we derive an explicit expression for the critical AMD below which the system is AMD stable.
Indeed, our construction of $\boldsymbol{e}_{12}$ from canonical variables through a rotation makes it particularly simple.
The total AMD $\equiv \mathcal{A}$ in the coplanar case is given by (Eqs.\:\ref{AMD}, \ref{canpoincare} and \ref{Gi})
\begin{equation}
\mathcal{A} = \sum_{i=1}^N \Gamma_i = \sum_{i=1}^N \frac{G_i^2}{2}
\end{equation}
where $N$ is the number of planets.
For any given pair of planets, which we label with indices 1 and 2, we can apply the canonical rotation $\mathcal{R}_\psi$ (Eq.\:\ref{Rpsibig}), with additional unity entries along the diagonals for the remaining planets, which remain unchanged.
The rotation preserves lengths, so we similarly have
\begin{equation}
\sum_{i=1}^N \frac{G_i^2}{2} = \sum_{i=1}^N \frac{F_i^2}{2} = \mathcal{A}.
\end{equation}

Finally, if we assume that all the AMD is transferred to $F_1$, which is simply a scaled version of $\tilde{e}_{12}$ , we obtain the maximum estimate of $\tilde{e}_{12}$ from Eq.\:\ref{F},
\begin{equation}
\tilde{e}_{12}^{max} = \sqrt{\frac{\Lambda_1 + \Lambda_2}{\Lambda_1 \Lambda_2}} \frac{\sqrt{2 \mathcal{A}}}{e_{12}^{\rm{cross}}} \label{emaxAMD}
\end{equation}

Combining this with Eq.\:\ref{H18crit}, we obtain the critical AMD below which the pair with indices 1 and 2 are AMD stable,
\begin{equation}
\mathcal{A}_{\rm{crit}} = \frac{\Lambda_1 \Lambda_2}{\Lambda_1 + \Lambda_2} \frac{{(e_{12}^{\rm{cross}}})^2}{2} \exp\Bigg[-\Bigg(\frac{3.0}{S_{12}}\Bigg)^{4/3}\Bigg] \label{AMDcrit}
\end{equation}

This procedure can be repeated for every adjacent pair of planets, and the minimum $\mathcal{A}_{\rm{crit}}$ then yields the critical AMD for the whole system.

Our expression is equivalent to that derived by \cite{Laskar17} in the compact limit ($\alpha \approx 1$), except for the additional exponential term in Eq.\:\ref{AMDcrit}.
Our corrected threshold can thus differ significantly from that of \cite{Laskar17} for systems with $S \lesssim 3$ (see Fig.\:\ref{fig:spacing}).

\bibliography{Bib}

\bibliographystyle{aasjournal}
\end{document}